\documentclass[journal]{IEEEtran}
\usepackage{graphicx,booktabs,array,multirow}
\usepackage{amsmath,amssymb,amsthm,cite}
\usepackage{xcolor,float,algpseudocode}
\usepackage[hidelinks]{hyperref}
\floatstyle{ruled}
\newfloat{algorithm}{tbp}{loa}
\floatname{algorithm}{Algorithm}
\algrenewcommand{\algorithmicrequire}{\textbf{Input:}}
\algrenewcommand{\algorithmicensure}{\textbf{Output:}}
\DeclareMathOperator*{\argmax}{arg\,max}

\newtheoremstyle{remarkstyle}%
  {}
  {}
  {\itshape}
  {}
  {\itshape}
  {.}
  {.5em}
  {}
\theoremstyle{remarkstyle}

\newif\ifshowresultfigures
\showresultfiguresfalse
\newcommand{\resultimage}[2]{%
\ifshowresultfigures
  \IfFileExists{#1}{\includegraphics[width=#2]{#1}}{\figureplaceholder{#1}{#2}}%
\else\figureplaceholder{#1}{#2}\fi}
\newcommand{\figureplaceholder}[2]{%
\fbox{\parbox[c][30mm][c]{\dimexpr#2-2\fboxsep-2\fboxrule\relax}{\centering\small Figure placeholder\\[4pt]\texttt{\detokenize{#1}}}}}

\newcommand{\conceptfigure}[3]{\begin{figure}[t]\centering
\IfFileExists{figures/#1.pdf}{\includegraphics[width=\columnwidth]{figures/#1.pdf}}{\figureplaceholder{figures/#1.pdf}{0.96\columnwidth}}
\caption{#2}\label{#3}\end{figure}}
\usepackage{tikz}
\usetikzlibrary{arrows.meta,positioning,calc}
\usepackage[caption=false,font=footnotesize]{subfig}   

\begin{document}
\title{Generative Beam Recovery via Joint Quality Modeling: Autoregression and Discrete Diffusion}\author{Amirhossein~Azarbahram,~\IEEEmembership{Member,~IEEE,}
        Onel~L.~A.~López,~\IEEEmembership{Senior~Member,~IEEE} 

\thanks{The authors are with Centre for Wireless Communications, University of Oulu, Finland, (e-mail: \{amirhossein.azarbahram, onel.alcarazlopez\}@oulu.fi). This work is supported by the Research Council of Finland (Grants 362782 (ECO-LITE), and 369116 (6G Flagship)).}%
}

\maketitle

\begin{abstract}
Millimeter-wave communication relies on directional
beamforming, requiring continuous beam management
to maintain alignment as propagation conditions change.
When the beam-quality feedback supporting this process
is interrupted, beam recovery must rely on outdated
observations and a limited number of beam probes. Multiple recovery outcomes may be consistent with the same observations, and ranking beams by their individual predicted quality can lead to redundant probing. To address this, we formulate beam recovery as a conditional maximum-coverage problem and develop two history-conditioned generative approaches based on autoregression and discrete diffusion. Both approaches generate joint beam-quality
profiles conditioned on pre-interruption feedback
and the elapsed time, which are used by a greedy
selection rule to choose probing
beams. Numerical results across multiple
deployments demonstrate that both approaches reduce probe
failure compared with categorical predictors and
generative baselines.
The differences between proposed models become more
pronounced when data availability is limited.
In particular, diffusion exhibits greater
robustness to sparse training labels and achieves
lower probe failure following limited
fine-tuning under tracker and mobility shifts.
This highlights the benefit of diffusion for beam recovery
under limited supervision and distribution shifts.
\end{abstract}

\IEEEpeerreviewmaketitle

\vspace{-3mm}
\section{Introduction}

\IEEEPARstart{M}{illimeter}-wave (mmWave) communication relies on directional transmission to compensate for severe propagation loss while exploiting large available bandwidths. However, narrow beams make link quality particularly sensitive to user mobility, orientation changes, and blockage, requiring timely beam management~\cite{mmwavesurvey2}.

In practical systems, beam management involves reference-signal transmission, reporting, and refinement \cite{giordaniBeamMgmt,ts38214}. Although exhaustive codebook sweeping provides great channel-quality information, it leads to significant signaling and measurement overhead \cite{rappapotmmwave,heath_mmwavemimo}. Consequently, beam decisions often rely on observations from a limited subset of directions. The difficulty increases when fresh observations become temporarily unavailable, e.g., due to interrupted measurements, delayed reporting, or a temporary loss of usable feedback. During such an interval, the user equipment (UE) may continue moving or changing orientation, leading to the serving beam being no longer suitable when measurements resume.

An observation gap introduces uncertainty about
the relative quality of alternative beams and the
optimum. Different motions and propagation
changes may be consistent with the same
pre-interruption history. This uncertainty
generally increases with the interruption
duration, as longer unobserved intervals allow
greater divergence in trajectories
and channel conditions. Moreover, each plausible
channel realization may admit several beams with
comparable quality, while beams that perform well
under one realization may be unsuitable for others.
Selecting beams solely by their individual predicted
quality may therefore yield redundant probes
that cover the same plausible outcomes.
Thus, under a strict probing budget, the challenge
is to identify a candidate subset likely to contain
sufficiently strong beams across plausible outcomes.  Restoring useful operation following a disturbance is an aspect of wireless resilience \cite{alves20256gresiliencewhite}, and in directional communication, efficient recovery probing can contribute to this objective.

\vspace{-3mm}
\subsection{Related work}

Learning-based beam prediction has extensively exploited auxiliary observations to reduce beam-search overhead. Situational awareness has been used to infer beam-related quantities in dynamic environments \cite{wang_situational_mmwave_beam_pred}, while sub-6~GHz channel state information (CSI) has been mapped to mmWave beam decisions \cite{alrabeiah_sub6_to_mmwave_beam_blockage} or combined with a limited number of mmWave pilot measurements \cite{dualb_fusion_hetnet_beam_pred}. Other low-overhead designs use learned radio features and temporal cross-band information \cite{lowcomplex_ml_mmwave_beam_pred,lstm_sub6_predictive_beam_tracking_v2i}. Visual, positional, and light detection and ranging (LiDAR) observations have also supported beam prediction \cite{vision_position_multimodal_beam_pred,lidar_future_beam_pred_v2i}, with Transformer-based methods integrating  sensing modalities \cite{multimodal_transformers_beam_pred}. The effectiveness of these approaches depends on the availability and reliability of the corresponding side information, which may be limited during observation interruptions.

Other studies reduce probing overhead by exploiting mmWave measurements or probabilistic information of beamspace. Learning-based initial-access methods infer suitable beams from a reduced set of measurements \cite{deepia_fast_initial_access}, while jointly learned probing patterns and prediction networks estimate a preferred beam from partial observations \cite{beam_probe_pattern_deep,site_specific_probe}. Beyond point prediction, Gaussian-process inference has been used to design limited beam-training subsets based on environmental priors \cite{zhang2023limitedsampling}. These establish the benefit of exploiting beam uncertainty and inter-beam structure when allocating a measurement budget. Their designs primarily concern beam acquisition or refinement using environmental priors or measurements collected during the probing procedure, not the initial probing decision following a period without observations.

Temporal beam prediction uses past observations to anticipate changes caused by mobility. Recurrent and sequential architectures have been investigated for beam prediction \cite{lowfreq_prior_lstm_beam_pred,dl_beam_tracking_under_mobility,multicell_multibeam_ae_lstm}. Time-domain reference signal received power (RSRP) has been used to forecast future RSRP and beam-switching within the 3rd Generation Partnership Project (3GPP) New Radio (NR) framework \cite{Jsac_time_mmwave_beam}. An ordinary differential equation (ODE)-long short-term memory (LSTM) architecture has also been used to forecast the preferred future beam based on historical full beam-training measurements~\cite{ode_lstm_continuous_time_beam_pred}. Predictive recovery under dynamic blockage, including alternative base station (BS) and beam selection, has also been studied \cite{bonfante2022predictiverecovery}. These works account for temporal evolution or link disruptions, but their targets and observations do not characterize the joint structure of codebook beams across multiple plausible post-interruption outcomes under a probing budget.

Generative modeling provides another means of representing uncertainty. In wireless systems, generative methods have been investigated for channel-related inference and semantic communication, including diffusion-based semantic reconstruction under ambiguity and channel perturbations \cite{Gen_wireless_1,Gen_wireless_2,pei2025latent,kim2026beyond}. More specifically, diffusion-based priors have been used to guide directional beam alignment from geometric and multipath information \cite{othman2026diffusionpriors}, while history-conditioned diffusion has been proposed to generate multiple future beamforming candidates using spatio-temporal priors \cite{wang2026beamstep}. Generative site-specific methods have also synthesized beamforming candidates from partial or decoupled channel measurements \cite{beam_brainstorm_genssbf,tang2026decoupledsensing}. Moreover, RadioDiff-v2 generates conditional angular power spectra from environmental geometry and employs sample-based greedy selection of complementary probing beams to improve the expected strongest received gain \cite{wang2026radiodiffv2}. 
These studies demonstrate the potential of
combining generative prediction with multi-beam
selection, although they differ in their
representations, conditioning information, and
selection objectives. However, exploiting the
joint distribution of codebook beam qualities
for limited-probing recovery after an observation
interruption remains insufficiently investigated.

\vspace{-4mm}
\subsection{Contributions}

\begin{figure}[t]
    \centering
    \resizebox{0.95\columnwidth}{!}{%
        \input{figs/fig_recovery_motivation_polished}%
    }
    \caption{The same feedback history may be consistent with a strong post-interruption beam in either the previously tracked sector or an alternative one. Coarse-to-fine search selects A and B by refining the tracked sector, individual ranking selects the same pair because they each cover more outcomes than C, and joint selection pairs A with C, covering two additional outcomes.}
    \label{fig:recovery_motivation}
\end{figure}



To address this gap, we consider beam recovery using only
pre-interruption feedback, where a limited probing set must
account for uncertainty in recovery-time beam qualities and
the possibility that multiple beams are acceptable under
the same channel realization. Existing beam-management procedures employ coarse-to-fine sweeping and refinement, using wider beams to identify promising directions before testing narrower beams~\cite{giordaniBeamMgmt,ts38214}. Learning-based recovery can also rank beams according to their individual predicted quality or acceptability. Under a limited probing budget, however, these may allocate multiple probes to beams that cover the same plausible post-interruption outcomes. Since only one beam is ultimately selected for serving, probing multiple beams that are acceptable under the same channel conditions provides no additional coverage, potentially leaving other plausible outcomes uncovered, as illustrated in Fig.~\ref{fig:recovery_motivation}. Our main contributions are:

\par\textbf{\textit{First}}, we formulate post-interruption beam candidate selection as a conditional coverage problem under a fixed probing budget. Accordingly, we seek to maximize the probability that at least one probed beam lies within a prescribed signal-to-noise ratio (SNR) gap of the full-information best beam. This formulation motivates learning the conditional joint distribution of beam-quality profiles to identify complementary candidates across uncertain recovery outcomes.

\par\textbf{\textit{Second}}, we develop two history-conditioned generative models for joint, discretized beam-quality profiles: i) an autoregressive model (ProfileAR) and ii) a discrete diffusion model (ProfileDiff). The former factorizes the profile distribution across beam coordinates, while the latter reconstructs masked quality levels through iterative refinement. During recovery, multiple profiles are sampled from the learned distributions, and a common greedy coverage rule selects probing beams to cover as many plausible outcomes as possible. This provides a unified framework for translating generative predictions into recovery decisions.

\par\textbf{\textit{Third}}, we reveal that:
\textit{(i)} joint generative modeling of beam-quality
profiles, combined with coverage-based selection,
improves recovery reliability over discriminative
and alternative generative baselines across
{the evaluated deployments;}
\textit{(ii)} autoregressive and diffusion models
achieve comparable recovery performance under
complete training supervision, highlighting the
benefits of joint profile modeling beyond the
choice of generative architecture;
\textit{(iii)} diffusion exhibits greater resilience
to incomplete supervision, with recovery performance
degrading more gradually as training labels become
sparse; and
\textit{(iv)} under tracker and mobility shifts,
diffusion achieves lower probe failure after
limited fine-tuning.

\textbf{Structure:} Section~\ref{sec:sysmodandprob} presents the system model and problem formulation. Section~\ref{sec:D3PM_OPT} introduces the generative models, followed by the coverage-based probing strategy in Section~\ref{sec:probing}. Sections~\ref{sec:sim_setup} and~\ref{sec:results} present the experimental setup and numerical results, respectively.

\textbf{Notations:} Bold lowercase and uppercase letters denote vectors and matrices, respectively. The Hermitian transpose is denoted by $(\cdot)^H$, and $\mathbf{1}\{\cdot\}$ denotes the indicator function. $\mathbb{E}[\cdot]$ and $\Pr(\cdot)$ denote expectation and probability, respectively, while $|\mathcal{A}|$ denotes the cardinality of a set $\mathcal{A}$. $\mathcal{N}(\mu,\sigma^2)$ denotes a real Gaussian distribution with mean $\mu$ and variance $\sigma^2$, and $\mathcal{CN}(0,\sigma^2)$ denotes a zero-mean circularly symmetric complex Gaussian distribution with variance $\sigma^2$. $\mathrm{Unif}(a,b)$ denotes the continuous uniform distribution over $[a,b]$.

\vspace{-2mm}
\section{System Model and Problem Formulation}
\label{sec:sysmodandprob}

We consider a downlink system comprising a BS equipped
with $N$ transmit antennas and a moving single-antenna
UE.\footnote{Coordinated multi-cell system is additionally investigated
in Sec.~\ref{sec:results}.} Time is divided into slots with equal duration,
indexed by $t$. The BS employs a finite beam codebook
$\mathcal{W}=\{\mathbf{w}_j\}_{j=0}^{K-1}$, where
$\mathbf{w}_j\in\mathbb{C}^{N}$ is a unit-norm
beamforming vector. The candidate beam space is
$\mathcal{K}=\{0,\ldots,K-1\}$.

Let $\mathbf{h}_t\in\mathbb{C}^{N}$ denote the downlink channel at slot $t$. When the BS
transmits using beam $j$ with power $P_{\mathrm{tx}}$,
the received signal is modeled as
\begin{equation}
r_{t,j}=
\sqrt{P_{\mathrm{tx}}}\,
\mathbf{h}_t^H\mathbf{w}_j x_t+n_t,
\label{eq:received_signal}
\end{equation}
where $x_t$ is a unit-power symbol and
$n_t\sim\mathcal{CN}(0,\sigma^2)$ is receiver noise.
The corresponding SNR is given by
\begin{equation}
\gamma_{t,j}=
{P_{\mathrm{tx}}
|\mathbf{h}_t^H\mathbf{w}_j|^2}/{\sigma^2},
\label{eq:snr_def}
\end{equation}
with dB value given by
$s_{t,j}=10\log_{10}(\gamma_{t,j})$.

The effective narrowband model allows us to focus on beam probing and selection, where wideband quality feedback may be represented by aggregated per-subcarrier SNRs. Frequency-dependent effects, such as beam squint~\cite{beam_squint_ref}, are not modeled. Moreover, the slot interval specifies the temporal sampling of channel realizations along the UE trajectory, capturing mobility-induced variations without assuming a constant channel within each slot or explicitly modeling Doppler effects.

\vspace{-4mm}
\subsection{Tracking history and interrupted observations}

As the UE moves, propagation conditions and
the preferred serving beam may change over time.
To maintain beam alignment, the BS employs a
tracking policy $\pi_{\mathrm b}$ that selects
one serving beam per slot based on the available
feedback history. Consistent with reference-signal-based
beam measurements (e.g., SSB/CSI-RS) and reporting supported by
3GPP NR~\cite{giordaniBeamMgmt,ts38214},
we consider UE feedback on the quality of the
selected serving beam, which informs subsequent
tracking decisions.


Suppose that this tracking process is interrupted,
e.g., due to blockage, missed feedback,
or temporary measurement unavailability.
Let $t_0$ denote the last slot with valid feedback
before the interruption. At $t_0$, the BS
has the served-beam indices and their reported
qualities over the past $L$ slots
\begin{equation}
\mathcal{H}=\{(b_\ell,y_\ell)\}_{\ell=1}^{L},
\qquad
y_\ell=Q_q(s_{t_\ell,b_\ell}+v_\ell),
\label{eq:history}
\end{equation}
where $t_\ell=t_0-L+\ell$,
$b_\ell\in\mathcal{K}$ is the serving-beam index,
and $v_\ell\sim\mathcal{N}(0,\sigma_v^2)$
models UE beam-quality measurement uncertainty
through a tractable Gaussian approximation in dB,
accounting for noisy, finite-duration observations
rather than channel variations, which are already
reflected in $s_{t_\ell,b_\ell}$.
The uniform quantizer $Q_q(\cdot)$, with step
size $q$ dB, represents finite-resolution
reporting of the measured beam quality.


During the interruption, the UE continues moving, but the BS receives no beam-quality update. Thus, the previously selected serving beam may no longer be suitable when measurements resume. After $\tau$ slot intervals, i.e., $t_{\mathrm r}=t_0+\tau$, the BS must identify suitable probing beams using the available history $\mathcal{H}$ and the known interruption duration $\tau$. Accordingly, we restrict the recovery input to $\mathcal{I}=(\mathcal{H},\tau)$, relying solely on beam-management feedback without assuming additional positioning or external sensing information. The interruption duration provides temporal context for predicting beam quality at recovery, as longer gaps allow greater divergence from observed channel conditions. Notably, $\tau$ can be determined at the BS from the timestamps of the last valid beam-quality report and the recovery event, without additional UE signaling.

\vspace{-4mm}
\subsection{Probe-and-serve recovery}

At recovery, the BS selects a subset of $P\ge 1$
beams for probing, denoted by
$\mathcal{P}\subseteq\mathcal{K}$, to limit
reference-signal measurement and feedback overhead.
The selected beams are measured within a recovery round, during which the channel is
assumed approximately constant.
For each $j\in\mathcal{P}$, the UE reports
$\widetilde y_j$, then the BS selects
the serving beam according to
\begin{equation}
\widehat j=
\argmax_{j\in\mathcal{P}}\widetilde y_j.
\label{eq:serve_rule}
\end{equation}
The resulting achieved SNR is
$\gamma_{t_{\mathrm r},\widehat j}$.

\begin{figure}[t]
    \centering
    \begin{tikzpicture}[x=1cm,y=1cm,>=Stealth,line join=round,line cap=round]
  \path[use as bounding box] (0,0) rectangle (8.50,5.12);

  \coordinate (beamApex) at (1.24,3.86);
  \path[fill=black!10]
    (beamApex)
      .. controls (1.85,4.08) and (2.65,4.34) .. (3.48,4.36)
      .. controls (3.70,4.36) and (3.72,4.28) .. (3.55,4.18)
      .. controls (2.80,4.11) and (2.02,3.99) .. (beamApex) -- cycle;
  \path[fill=black!20]
    (beamApex)
      .. controls (1.95,3.97) and (2.90,4.05) .. (3.82,4.02)
      .. controls (4.05,4.01) and (4.05,3.92) .. (3.86,3.83)
      .. controls (3.01,3.80) and (2.05,3.80) .. (beamApex) -- cycle;
  \path[fill=black!10]
    (beamApex)
      .. controls (1.88,3.69) and (2.64,3.45) .. (3.38,3.38)
      .. controls (3.62,3.36) and (3.63,3.28) .. (3.46,3.20)
      .. controls (2.79,3.25) and (2.00,3.48) .. (beamApex) -- cycle;

  \begin{scope}[shift={(.65,3.84)},scale=.65]
    \draw[line width=1.55pt] (-.37,-.55)--(0,.45)--(.37,-.55)--cycle;
    \draw[line width=.60pt] (-.23,-.13)--(.23,-.13)
                            (-.29,-.36)--(.29,-.36);
    \fill (0,.45) circle (.09);
    \draw[line width=.7pt] (-.17,.71) arc (135:225:.30);
    \draw[line width=.7pt] ( .17,.71) arc (45:-45:.30);
    \draw[line width=.7pt] (-.32,.94) arc (135:225:.61);
    \draw[line width=.7pt] ( .32,.94) arc (45:-45:.61);
  \end{scope}
  \node[font=\sffamily\scriptsize\bfseries] at (.65,3.22) {BS};

  \node[anchor=west,font=\sffamily\scriptsize] at (1.48,4.92)
    {Codebook $\mathcal{W}=\{\mathbf{w}_j\}_{j=0}^{K-1}$};
  \draw[thin,-{Stealth[length=1.2mm]}] (2.54,4.75)--(2.35,4.43);

  \newcommand{\drawcar}[2]{%
    \begin{scope}[shift={(#1,#2)}]
      \path[fill=black]
        (-.38,-.10)--(-.35,.05)--(-.20,.10)--(-.09,.24)
        --(.13,.24)--(.27,.07)--(.37,.02)--(.40,-.10)--cycle;
      \path[fill=white]
        (-.16,.10)--(-.06,.21)--(.02,.21)--(.02,.10)--cycle;
      \path[fill=white]
        (.07,.10)--(.07,.21)--(.13,.21)--(.24,.10)--cycle;
      \foreach \xx in {-.22,.25}{
        \fill[white] (\xx,-.10) circle (.080);
        \fill[black] (\xx,-.10) circle (.052);
      }
    \end{scope}%
  }
  \drawcar{3.60}{3.77}
  \drawcar{7.43}{3.77}
  \node[font=\footnotesize] at (3.60,3.25) {$t_0$};
  \node[font=\footnotesize] at (7.42,3.25)
    {$t_{\mathrm r}=t_0+\tau$};

  \draw[densely dotted,line width=.70pt]
    (4.05,3.75).. controls (4.60,3.81) and (4.83,3.74).. (5.20,3.70);
  \draw[densely dotted,line width=.70pt,-{Stealth[length=1.4mm]}]
    (6.07,3.70).. controls (6.57,3.71) and (6.91,3.75).. (7.01,3.79);

  \begin{scope}[shift={(5.60,3.48)}]
    \fill[black!45] (-.39,0) rectangle (-.13,.48);
    \fill[black!60] (-.09,0) rectangle (.24,.76);
    \fill[black!45] (.28,0) rectangle (.52,.48);
    \foreach \x in {-.31,-.21}{
      \foreach \y in {.12,.28}{
        \fill[white] (\x,\y) rectangle ++(.055,.075);
      }
    }
    \foreach \x in {-.01,.12}{
      \foreach \y in {.14,.34,.54}{
        \fill[white] (\x,\y) rectangle ++(.065,.08);
      }
    }
    \foreach \x in {.34,.43}{
      \foreach \y in {.12,.29}{
        \fill[white] (\x,\y) rectangle ++(.05,.065);
      }
    }
  \end{scope}

  \draw[thin] (3.63,4.29)--(3.63,4.46) (7.40,4.29)--(7.40,4.46);
  \draw[thin,<->] (3.70,4.38)--(7.33,4.38);
  \node[fill=white,inner sep=1.8pt,font=\sffamily\scriptsize]
    at (5.52,4.60) {Interruption ($\tau$ slots)};

  \draw[black!60,line width=.5pt] (.16,2.92)--(8.34,2.92);
  \node[font=\sffamily\bfseries\footnotesize] at (4.25,2.65)
    {Recovery event at $t_{\mathrm r}$};

  \tikzset{
    stepbox/.style={
      draw=black,line width=.48pt,rounded corners=.8pt,
      minimum width=3.67cm,minimum height=.85cm,inner sep=0pt
    }
  }
  \node[stepbox] (A) at (2.16,1.87) {};
  \node[stepbox] (B) at (6.34,1.87) {};
  \node[stepbox] (C) at (6.34,.70) {};
  \node[stepbox] (D) at (2.16,.70) {};

  \node[font=\sffamily\bfseries\scriptsize] at ($(A.center)+(0,.16)$)
    {History and outage duration};
  \node[font=\footnotesize] at ($(A.center)+(0,-.18)$)
    {$\mathcal{I}=(\mathcal{H},\tau)$};

  \node[font=\sffamily\bfseries\scriptsize] at ($(B.center)+(0,.16)$)
    {Select probing set};
  \node[font=\footnotesize] at ($(B.center)+(0,-.18)$)
    {$\mathcal{P}\subseteq\mathcal{K},\quad |\mathcal{P}|=P$};

  \node[font=\sffamily\bfseries\scriptsize] at ($(C.center)+(0,.16)$)
    {Probe and obtain feedback};
  \node[font=\footnotesize] at ($(C.center)+(0,-.18)$)
    {$\{\widetilde y_j\}_{j\in\mathcal{P}}$};

  \node[font=\sffamily\bfseries\scriptsize] at ($(D.center)+(0,.16)$)
    {Serve best reported beam};
  \node[font=\footnotesize] at ($(D.center)+(0,-.18)$)
    {$\widehat j=\argmax_{j\in\mathcal{P}}\widetilde y_j$};

  \draw[thin,-{Stealth[length=1.35mm]}]
    ($(A.east)+(.03,0)$)--($(B.west)+(-.03,0)$);
  \draw[thin,-{Stealth[length=1.35mm]}]
    ($(B.south)+(0,-.03)$)--($(C.north)+(0,.03)$);
  \draw[thin,-{Stealth[length=1.35mm]}]
    ($(C.west)+(-.03,0)$)--($(D.east)+(.03,0)$);
\end{tikzpicture}
    \caption{System model for codebook-based beam
    recovery after an observation interruption
    of duration $\tau$.}
    \label{fig:system_model}
\end{figure}

\vspace{-4mm}
\subsection{Problem formulation}

The goal is to select $\mathcal{P}$ using the
available information $\mathcal{I}$ without
observing the recovery channel. Since individual
beam preferences may lead to redundant probes,
we formulate this as a conditional coverage
problem that accounts for complementary beams
under a fixed probing budget. Let us define the full-information oracle beam quality
at recovery as
$s_{t_{\mathrm r}}^\star
=\max_{j\in\mathcal{K}}s_{t_{\mathrm r},j}$.
For a predefined acceptable quality loss
$\Delta_{\mathrm{acc}}\geq 0$ in dB, the set of
acceptable beams is 
\begin{equation}
\mathcal{A}_{t_{\mathrm r}}
=\left\{j\in\mathcal{K}:
s_{t_{\mathrm r}}^\star-s_{t_{\mathrm r},j}
\leq\Delta_{\mathrm{acc}}\right\}.
\label{eq:acceptable}
\end{equation}
A probing set is considered successful
if it contains at least one acceptable beam.
Accordingly, the probe-failure indicator is $F_{\mathrm p}(\mathcal{P})
=\mathbf{1}\left\{
\mathcal{P}\cap\mathcal{A}_{t_{\mathrm r}}
=\varnothing\right\}$. Let $\mu$ denote a recovery policy that maps the
available information $\mathcal{I}$ to a probing
set $\mathcal{P}$. Our objective is to minimize
the expected probe-failure probability subject
to a fixed probing budget, formulated as
\begin{subequations}\label{eq:primary_problem}
\begin{align}
\underset{\mu}{\operatorname{minimize}}\quad&
\mathbb{E}[F_{\mathrm p}(\mu(\mathcal{I}))]
\label{eq:primary_objective}\\
\text{subject to}\quad&
\mathcal{P}=\mu(\mathcal{I})\subseteq\mathcal{K},
\quad |\mathcal{P}|=P.
\label{eq:budget_constraint}
\end{align}
\end{subequations}
The probing action space contains
$\binom{K}{P}$ possible subsets, making
exhaustive subset selection computationally
demanding for large codebooks. Moreover, the expectation is  w.r.t. recovery events,
including their observation histories, interruption
durations, and channel realizations at recovery.
For a given $\mathcal{I}$, the history and
interruption duration are known, while the
recovery beam qualities remain uncertain.
By the law of total expectation, and since
probing decisions are not coupled across recovery
events, minimizing the expected probe-failure
probability is equivalent to maximizing the
conditional probability of probe success for
each $\mathcal{I}$, yielding
\begin{equation}
\mu^{\star}(\mathcal{I})
\in \arg\max_{\mathcal{P}\subseteq\mathcal{K},\,|\mathcal{P}|=P}
\Pr\left\{
\mathcal{P}\cap\mathcal{A}_{t_{\mathrm r}}
\neq\varnothing\mid\mathcal{I}
\right\}.
\label{eq:coverage_objective}
\end{equation}

This leads to a conditional stochastic
maximum-coverage problem, whose evaluation
requires the conditional joint distribution
of beam acceptability. We therefore develop
history-conditioned generative models of
future beam-quality profiles, whose samples
enable greedy approximation of
\eqref{eq:coverage_objective} without exhaustive
enumeration of probing sets. Note that for $P=1$, selecting the beam with the highest
conditional acceptability probability is optimal.
For $P>1$, however, beams with high marginal probabilities may cover the same
channel realizations, leading to redundant probes.
Thus, optimal probing-set selection
requires accounting for joint structure.

\vspace{-2mm}
\section{Generative Beam-Quality Modeling}
\label{sec:D3PM_OPT}

Evaluating the coverage objective in
\eqref{eq:coverage_objective} requires
characterizing which beams can be acceptable
together under plausible recovery outcomes.
Three natural target representations are:
i) binary acceptability profiles, which directly
capture the coverage events but discard
information about relative beam-quality losses;
ii) continuous-valued profiles, which preserve
the full quality information but require modeling
fine-scale variations that are not necessary
for the prescribed coverage criterion; and
iii) discretized quality-gap profiles, which
preserve beam acceptability exactly while also
retaining coarse information about relative
losses. We adopt the third representation, as it
preserves decision-relevant beam-quality
information beyond binary acceptability
without requiring the fine-grained distribution
modeling associated with continuous-valued
profiles. We then learn the conditional joint distribution
of discretized beam-quality profiles given
$\mathcal{I}$ using two approaches:
autoregressive modeling and discrete denoising
diffusion. Both approaches share the same
target representation and conditioning
information, but differ in how they capture
and generate dependencies among beam qualities.

\vspace{-4mm}
\subsection{Discretized beam-quality representation}
\label{subsec:trainopt}

To represent the candidate beams' qualities at
recovery, we first define the gap between each beam
and the oracle as
\begin{equation}
g_j=s_{t_{\mathrm r}}^\star-s_{t_{\mathrm r},j},
\label{eq:quality_gap}
\end{equation}
where $g_j\geq 0$ and smaller values indicate
better beams. We quantize the beam-quality gaps into $Q$ ordered levels using $Q-1$ thresholds,
$0\leq\delta_1=\Delta_{\mathrm{acc}}
<\delta_2<\cdots<\delta_{Q-1}$.
The first level represents acceptable beams with quality gaps not exceeding $\Delta_{\mathrm{acc}}$, while subsequent levels represent progressively larger losses. The discrete beam-quality
profile $\mathbf{x}_0\in\{0,\ldots,Q-1\}^K$
is defined elementwise as
\begin{equation}
x_{0,j}=\sum_{a=1}^{Q-1}\mathbf{1}\{g_j>\delta_a\}.
\label{eq:profile}
\end{equation}
Hence,
$x_{0,j}=0$ identifies an acceptable beam,
i.e., $j\in\mathcal{A}_{t_{\mathrm r}}$.

Unlike predicting a single optimal index,
$\mathbf{x}_0$ characterizes the quality of
all $K$ beams. In particular,
it identifies multiple acceptable alternatives
and retains coarse quality-gap information beyond binary acceptability. Learning its joint distribution allows
the model to capture dependencies among beam
qualities and the uncertainty in their future
realizations. We therefore aim to approximate
the distribution $p(\mathbf{x}_0\mid\mathcal{I})$.

\vspace{-4mm}
\subsection{History encoder}
\label{subsec:history_encoder}

To predict recovery-time beam qualities, we must extract relevant information from historical beam measurements while accounting for their temporal evolution and the elapsed interruption duration. To this end, we employ a shared Transformer-based history encoder that maps the available information $\mathcal{I}$ into a contextual representation to condition profile generation.

For each historical observation
$(b_\ell,y_\ell)\in\mathcal{H}$, we construct
a token combining the serving-beam identity,
reported quality, feedback variation, and
temporal position such that
\begin{align}
\mathbf{e}_\ell
&=\operatorname{BeamTok}(b_\ell)
+\operatorname{MLP}(\mathbf{z}_\ell)
+\mathbf{p}_\ell,\\
\mathbf{z}_\ell
&=\left[
{(y_\ell-\mu_y)}/{\sigma_y},
{\Delta y_\ell}/{\sigma_{\Delta y}}
\right]^T,
\label{eq:history_token}
\end{align}
where $\Delta y_\ell=y_\ell-y_{\ell-1}$
for $\ell>1$, and $\Delta y_1=0$.
The parameters $\mu_y$, $\sigma_y$, and
$\sigma_{\Delta y}$ are normalization
statistics of the training data,
while $\mathbf{p}_\ell$ is a temporal
positional embedding. The beam tokenizer combines a learned beam-index embedding with fixed features encoding the corresponding beam direction. Similarly, we encode $\tau$ with sinusoidal features and map it through a multi-layer perceptron (MLP) to obtain a duration token $\mathbf{e}_\tau$. Then, 
$\{\mathbf{e}_\ell\}_{\ell=1}^{L}$
and $\mathbf{e}_\tau$
are concatenated and processed by a
Transformer encoder with $N_{\mathrm{enc}}$
layers producing
the contextual memory $\mathbf{M}=f_\phi(\mathcal{I})$, where $\phi$ collects the encoder parameters, as summarized in Fig.~\ref{fig:encoder}. 

\begin{figure}[t]
\centering
\begin{tikzpicture}[
  x=1cm,y=1cm,
  font=\sffamily\footnotesize,
  >={Latex[length=1.65mm,width=1.08mm]},
  flow/.style={->,line width=.65pt,draw=black!70},
  input/.style={draw=black!65,fill=white,rounded corners=2pt,
      line width=.55pt,align=center,inner xsep=2pt,inner ysep=4pt},
  process/.style={draw=black!65,fill=black!3,rounded corners=2pt,
      line width=.6pt,align=center,inner xsep=2pt,inner ysep=4pt},
  encoder/.style={draw=black!80,fill=blue!9,rounded corners=2pt,
      line width=.75pt,align=center,inner xsep=3pt,inner ysep=4pt}
]
\def\leftX{-1.36}
\def\rightX{2.71}
\def\inputY{1.55}
\def\processY{0.20}
\def\encY{-1.32}

\node[input,minimum width=5.20cm,minimum height=.66cm]
  (history) at (\leftX,\inputY)
  {\scriptsize \textbf{Tracking history}\quad$\mathcal H=\{(b_\ell,y_\ell)\}_{\ell=1}^{L}$};
\node[input,minimum width=2.50cm,minimum height=.66cm]
  (duration) at (\rightX,\inputY)
  {\scriptsize \textbf{Duration}\quad$\tau$};

\node[process,minimum width=5.20cm,minimum height=1.06cm]
  (historytoken) at (\leftX,\processY)
  {\scriptsize \textbf{History tokenization}\\[1pt]
   {\scriptsize $\mathbf e_\ell=\operatorname{BeamTok}(b_\ell)+\operatorname{MLP}(\mathbf z_\ell)+\mathbf p_\ell$}\\[1pt]
   {\scriptsize $\mathbf z_\ell$: normalized $y_\ell$ and $\Delta y_\ell$}};
\node[process,minimum width=2.50cm,minimum height=1.06cm]
  (durationtoken) at (\rightX,\processY)
  {\scriptsize \textbf{Duration token}\\[2pt]
   {\scriptsize sinusoid$(\tau)$ $\rightarrow$ MLP}};

\node[encoder,minimum width=7.92cm,minimum height=0.75cm]
  (transformer) at (0,\encY)
  {\scriptsize \textbf{Transformer history encoder}\\[-1pt]
   $N_{\mathrm{enc}}$ layers \quad $\mathbf M=f_\phi(\mathcal I)$};
\draw[flow] (history.south) -- (historytoken.north);
\draw[flow] (duration.south) -- (durationtoken.north);
\draw[flow] (historytoken.south) -- ([xshift=-1.36cm]transformer.north);
\draw[flow] (durationtoken.south) -- ([xshift=2.71cm]transformer.north);
\end{tikzpicture}
\caption{The history-encoder architecture.}
\label{fig:encoder}
\end{figure}

\vspace{-4mm}
\subsection{{ProfileAR}: autoregressive generation}
\label{subsec:ar_profile}

The first approach models the conditional distribution
$p(\mathbf{x}_0\mid\mathcal{I})$ using autoregressive
generation, which represents a joint distribution
as a product of conditional distributions
\cite{parmar2018image}. 
The joint distribution of the quality levels
is factorized as
\begin{equation}
p_\theta^{\mathrm{AR}}(\mathbf{x}_0\mid\mathcal{I})
=
\prod_{j=0}^{K-1}
p_\theta^{\mathrm{AR}}
(x_{0,j}\mid\mathbf{x}_{0,<j},\mathbf{M}),
\label{eq:ar_factorization}
\end{equation}
where $\mathbf{x}_{0,<j}=
(x_{0,0},\ldots,x_{0,j-1})$ denotes the
quality levels preceding beam $j$ in the
fixed codebook ordering, with an empty
sequence for $j=0$, and $\theta$
denotes the generator parameters. We implement the conditional distributions using a Transformer decoder~\cite{vaswani2017attention}. Causal self-attention captures dependencies among beam-quality levels, while cross-attention incorporates the history memory $\mathbf{M}$. Each output position corresponds to a fixed codebook beam and predicts a categorical distribution over its $Q$ quality levels.

During training, we employ teacher forcing, where the decoder receives the ground-truth quality levels shifted by one position. Causal self-attention masking ensures that the prediction at beam coordinate $j$ depends only on preceding quality levels, preventing access to the current or subsequent targets. This enables all conditional distributions to be evaluated in parallel during training while preserving the autoregressive factorization. For a minibatch $\mathcal{D}_{\mathrm{mb}}$ of training pairs, the encoder and decoder parameters are jointly optimized by minimizing the negative log-likelihood given by
\begin{equation}
\widehat{\mathcal{L}}_{\mathrm{AR}}
\!=\!{}-\frac{1}{|\mathcal{D}_{\mathrm{mb}}|K}
\!\sum_{r\in\mathcal{D}_{\mathrm{mb}}}\!
\!\sum_{j=0}^{K-1}\!
\log p_\theta^{\mathrm{AR}}
\left(x_{0,j}^{(r)}
\mid\mathbf{x}_{0,<j}^{(r)},
f_\phi(\mathcal{I}_r)\right).
\label{eq:ar_loss}
\end{equation}
At inference, the ground truth is unavailable. The decoder starts from a
dedicated start token and sequentially samples the
quality level of beams, using the
previously generated levels as context.
Sampling probabilities are obtained by
dividing the predicted logits by a
temperature $T_s>0$ before softmax.
After $K$ steps, the generated sequence
forms a complete profile
$\mathbf{x}_0$.

\vspace{-4mm}
\subsection{{ProfileDiff}: discrete diffusion-based generation}
\label{subsec:diff_profile}

The second approach models
$p(\mathbf{x}_0|\mathcal{I})$ using
discrete denoising diffusion \cite{D3PM_main}. Specifically, it constructs a complete
profile by progressively revealing masked
coordinates. We adopt an absorbing-mask
diffusion process
\cite{shi2024masked}, comprising
a forward corruption process and a
conditional reverse process.

\subsubsection{Forward corruption process}

We augment the quality-level alphabet
$\{0,\ldots,Q-1\}$ with a mask symbol
$\mathsf{u}$. Let $\xi\in[0,1]$ denote
diffusion time, distinct from $t$ and $\tau$. Given a clean profile $\mathbf{x}_0$,
each beam coordinate is independently
masked with probability $\xi$ such that
\begin{align}
q(x_{\xi,j}\mid x_{0,j})
=(1-\xi)\mathbf{1}\{x_{\xi,j}=x_{0,j}\}
+\xi\mathbf{1}\{x_{\xi,j}=\mathsf{u}\}.
\label{eq:mask_forward}
\end{align}
Thus, $\xi=0$ corresponds to the clean
profile, while $\xi=1$ produces a fully
masked profile. Once a coordinate is masked, it
remains masked as diffusion time increases.

\subsubsection{Conditional denoising and training}

A Transformer denoiser is trained to predict
the clean quality levels from a corrupted
profile and the history-encoder memory
$\mathbf{M}=f_\phi(\mathcal{I})$.
Each beam token combines its identity,
corrupted quality level, and diffusion-time
embedding. Self-attention allows the
denoiser to exploit dependencies among
beam coordinates, while cross-attention
incorporates the historical information.

For each beam $j$, the denoiser outputs
a categorical distribution over the
$Q$ clean quality levels denoted by
\begin{equation}
\boldsymbol{\pi}_{\theta,j}
(\mathbf{x}_\xi,\xi,\mathbf{M})
\in[0,1]^Q,
\quad
\sum_{a=0}^{Q-1}\pi_{\theta,j,a}=1,
\label{eq:denoiser_output}
\end{equation}
where $\pi_{\theta,j,a}$ is the
predicted probability that the clean
quality level of beam $j$ equals $a$.

During training, a diffusion time
$\xi_r\sim\mathrm{Unif}(\epsilon,1)$,
with a small $\epsilon>0$, is sampled
for each training pair $r$. The clean
profile $\mathbf{x}_0^{(r)}$ is then
corrupted according to
\eqref{eq:mask_forward}. The denoiser
predicts the original quality levels
at the masked coordinates, using the
unmasked coordinates as additional
context.

Let us define
the set of masked coordinates as
\begin{equation}
\mathcal{V}=
\left\{(r,j):
r\in\mathcal{D}_{\mathrm{mb}},
j\in\mathcal{K},
x_{\xi_r,j}^{(r)}=\mathsf{u}
\right\}.
\label{eq:masked_coordinates}
\end{equation}
The history encoder and denoiser are
jointly trained by minimizing the
average cross-entropy over these
coordinates as
\begin{align}
\widehat{\mathcal{L}}_{\mathrm{Diff}}
=-\frac{1}{|\mathcal{V}|}
\sum_{(r,j)\in\mathcal{V}}
\log\pi_{\theta,j,x_{0,j}^{(r)}}
\left(
\mathbf{x}_{\xi_r}^{(r)},\xi_r,
f_\phi(\mathcal{I}_r)
\right).
\label{eq:train_loss}
\end{align}
The loss is evaluated only at masked
coordinates, whose original levels
are known from the clean training
profiles. Minibatches with $\mathcal{V}=\varnothing$ are skipped.

\subsubsection{Reverse generation}

At inference, generation begins with
a fully masked profile
$\mathbf{x}_{\xi_0}=\mathsf{u}^K$.
We consider a decreasing diffusion-time
grid
$1=\xi_0>\xi_1>\cdots>\xi_T=0$,
where $T$ denotes the number of reverse
steps. At each step $i$, the denoiser predicts
the clean-level distributions from
$\mathbf{x}_{\xi_{i-1}}$,
$\xi_{i-1}$, and $\mathbf{M}$.
Previously revealed quality levels
remain unchanged. Each coordinate
that is still masked either remains
masked or is assigned a quality
sampled from the predicted distribution. For
$x_{\xi_{i-1},j}=\mathsf{u}$,
the reverse transition is given by
\begin{subequations}\label{eq:reverse_transition}
\begin{align}
&p_\theta\left(
x_{\xi_i,j}=\mathsf{u}
\mid\mathbf{x}_{\xi_{i-1}},\mathcal{I}
\right)
={\xi_i}/{\xi_{i-1}},
\label{eq:reverse_transition_a}
\\
&p_\theta\left(
x_{\xi_i,j}=a
\mid\mathbf{x}_{\xi_{i-1}},\mathcal{I}
\right)
=
\left(1-{\xi_i}/{\xi_{i-1}}\right)
\pi_{\theta,j,a}^{(T_s)},
\label{eq:reverse_transition_b}
\end{align}
\end{subequations}
where $a\in\{0,\ldots,Q-1\}$ and
$\pi_{\theta,j,a}^{(T_s)}$ is obtained
by applying softmax to the denoiser
logits divided by the sampling
temperature $T_s>0$. Since $\xi_T=0$, all remaining masked
coordinates are revealed at the final
step, yielding a complete
profile $\mathbf{x}_0$.

Although beam-quality levels revealed within the same denoising step are sampled conditionally independently, subsequent steps condition on previously revealed levels. This allows the model to capture cross-beam dependencies without imposing a fixed generation order. However, simultaneously revealed levels are not sampled from their exact joint conditional distribution, introducing an approximation in finite-step generation. In particular, $T=1$ reduces to independent categorical sampling conditioned only on the history.

\vspace{-4mm}
\subsection{Offline training and supervision}

\begin{algorithm}[t]
\caption{Offline training of the profile generators}
\label{alg:training}
\begin{algorithmic}[1]
\Require Training/validation events with labeled
coordinates $\mathcal{O}_r$; generator
$\mathrm{G}\in\{\mathrm{ProfileAR},\mathrm{ProfileDiff}\}$;
iteration budget $N_{\mathrm{it}}$;
validation period $N_{\mathrm{v}}$;
patience $N_{\mathrm{p}}$
\Ensure Trained encoder $f_\phi$, generator $p_\theta$,
and $T_s$
\State Quantize gaps relative to the strongest measured beam in each future using \eqref{eq:profile}
\For{$i=1,\dots,N_{\mathrm{it}}$}
    \State Sample a minibatch of events and
    one future per event
    \State Compute $\mathbf{M}=f_\phi(\mathcal{I})$
    \If{$\mathrm{G}=\mathrm{ProfileAR}$}
        \State Form teacher-forced inputs,
        replacing missing preceding levels
        with an unknown token
        \label{line:ar_start}
        \State Compute $\mathcal{L}$
        by \eqref{eq:ar_loss} over labeled
        targets only
        \label{line:ar_end}
    \Else\ sample $\xi_r\sim
        \mathrm{Unif}(\epsilon,1)$ per event
        \label{line:diff_start}, corrupt labeled coordinates by
        \eqref{eq:mask_forward}, and
        keep unlabeled ones masked
        \State Exclude samples $r$ for which
        $\mathcal{M}_{\xi_r}\cap\mathcal{O}_r=\varnothing$
        \If{no samples remain}\ \textbf{continue}
        \EndIf
        \State Compute $\mathcal{L}$ by \eqref{eq:partial_loss} over the remaining samples
        \label{line:diff_end}
    \EndIf
    \State Update $(\phi,\theta)$ by minimizing
    $\mathcal{L}$ using AdamW 
    and a warm-up/cosine learning-rate schedule
    \If{$i\bmod N_{\mathrm{v}}=0$}
        \State Evaluate validation loss
        and save best parameters
        \If{no improvement for
        $N_{\mathrm{p}}$ evaluations}\ \textbf{break}
        \EndIf
    \EndIf
\EndFor
\State Restore the best $(\phi,\theta)$
\State Select $T_s$ minimizing validation
probe-failure probability
\end{algorithmic}
\end{algorithm}

Both generative models use discretized joint
beam-quality profiles as training targets.
For complete profiles, the quality gaps are computed
relative to the strongest beam in the full codebook.
Offline training profiles can be obtained
from ray-traced channel simulations,
dedicated full-codebook measurement campaigns,
or channel-sounding procedures that provide
sufficient channel information to reconstruct
the codebook beam qualities. A complete training
profile requires the {qualities of all $K$ beams.}
Acquiring these profiles in real deployments
can incur substantial measurement overhead
and requires sufficiently stable channel
conditions during acquisition. Consequently,
practical training datasets may contain
missing beam-quality labels. We therefore
consider both complete and partially labeled
training profiles, with the latter requiring
a suitable reference for constructing the
available quality-gap labels. Such supervision
is used exclusively during offline training
and is unavailable during recovery.
Algorithm~\ref{alg:training} summarizes the
training procedure, with
lines~\ref{line:ar_start}--\ref{line:ar_end}
and \ref{line:diff_start}--\ref{line:diff_end}
corresponding to ProfileAR and ProfileDiff,
respectively.

Since complete beam-quality measurements may not always
be available during offline data collection, we also consider
training with incomplete profiles.
Let $\mathcal O\subseteq\mathcal K$ denote the set of beams
with available quality measurements. For each $j\in\mathcal O$,
we quantize the gap $\max_{k\in\mathcal O}s_{t_{\mathrm r},k}
-s_{t_{\mathrm r},j}$ using the thresholds in Sec.~III-A.
Thus, level zero denotes a beam within $\Delta_{\mathrm{acc}}$
of the strongest measured beam.
For ProfileAR, missing levels are replaced
by a dedicated unknown token in the input sequence, while
the loss is evaluated over labeled coordinates.
Notably, this introduces a
mismatch between training, where unknown
tokens may appear in conditioning prefixes,
and inference, where previously generated
levels are used.

For ProfileDiff, unlabeled coordinates remain
masked, while the known coordinates are
randomly corrupted according to
\eqref{eq:mask_forward}. The reconstruction
loss is evaluated only at masked coordinates
with known labels, yielding
\begin{equation}
\mathcal{L}_{\mathcal{O}}
=-\mathbb{E}\left[
\frac{1}{|\mathcal{M}_\xi\cap\mathcal{O}|}
\sum_{j\in\mathcal{M}_\xi\cap\mathcal{O}}
\log p_\theta(x_{0,j}\mid
\mathbf{x}_\xi,\xi,\mathbf{M})
\right],
\label{eq:partial_loss}
\end{equation}
where $\mathcal{M}_\xi$ denotes the masked
coordinates, and samples with
$\mathcal{M}_\xi\cap\mathcal{O}=\varnothing$
are excluded. Unlike ProfileAR, ProfileDiff
can incorporate incomplete profiles through its masking mechanism without introducing
additional input symbols or imputing missing
labels, which is
beneficial under sparse supervision.

\section{Coverage-Based Probing and Deployment}
\label{sec:probing}


Given $\mathcal{I}$, trained generators produce $S$
independent profile samples
$\{\mathbf{x}_0^{(r)}\}_{r=1}^{S}$.
Each sample is converted into a predicted
acceptable-beam set as
\begin{equation}
\widehat{\mathcal{A}}_r
=
\{j\in\mathcal{K}:x_{0,j}^{(r)}=0\},
\qquad r=1,\ldots,S.
\label{eq:generated_sets}
\end{equation}
The generated profiles separate uncertainty modeling from the probing decision. A single set of samples can support different budgets $P$ through the coverage objective explained below.

Using the sampled sets in \eqref{eq:generated_sets}, we estimate the probability that a probing set covers at least one acceptable beam.
For $\mathcal{P}\subseteq\mathcal{K}$, its empirical coverage is given by
\begin{equation}
\widehat{\mathcal{C}}(\mathcal{P})
=
\frac{1}{S}
\sum_{r=1}^{S}
\mathbf{1}
\left\{
\mathcal{P}\cap
\widehat{\mathcal{A}}_r
\neq\varnothing
\right\}.
\label{eq:empirical_coverage}
\end{equation}
This objective accounts
for overlap among their acceptable
sets across sampled futures.

\vspace{-4mm}
\subsection{Greedy probing-set selection}

Given the generated acceptable-beam sets, we seek $P$ probing beams that cover as many sampled recovery outcomes as possible. We adopt a greedy strategy that iteratively selects the beam providing the largest additional coverage. Starting from $\mathcal{P}_0=\varnothing$, the marginal gain of an unselected beam $j$ at iteration $i$ is the number of sampled outcomes in which $j$ is acceptable, but none of the previously selected beams, as
\begin{equation}
d_i(j)=
\sum_{r=1}^{S}
\mathbf{1}\{j\in\widehat{\mathcal{A}}_r\}
\mathbf{1}\left\{
\mathcal{P}_{i-1}\cap
\widehat{\mathcal{A}}_r
=\varnothing
\right\}.
\label{eq:greedy_gain}
\end{equation}
The beam with the largest marginal gain is added to the probing set such that
\begin{equation}
j_i\in
\arg\max_{j\in
\mathcal{K}\setminus\mathcal{P}_{i-1}}
d_i(j),
\qquad
\mathcal{P}_i=
\mathcal{P}_{i-1}\cup\{j_i\}.
\label{eq:greedy_selection}
\end{equation}
This process continues until $P$ distinct beams are selected, while ties are resolved by sample frequency and index order.

The empirical coverage function in
\eqref{eq:empirical_coverage} is monotone
and submodular, as adding beams cannot
reduce coverage and exhibits diminishing
marginal gains. Consequently, the greedy
solution $\mathcal{P}_{\mathrm g}=\mathcal{P}_P$
satisfies~\cite{nemhauser1978analysis}
\begin{equation}
\widehat{\mathcal{C}}(\mathcal{P}_{\mathrm g})
\geq
\left[1-\left(1-\frac{1}{P}\right)^P\right]
\max_{|\mathcal{P}|=P}
\widehat{\mathcal{C}}(\mathcal{P}).
\label{eq:greedy_bound}
\end{equation}
This guarantee concerns empirical coverage
over generated samples, not the true recovery
probability. Fig.~\ref{fig:online_workflow}
and Algorithm~\ref{alg:online} summarize
the inference procedure. Lines~\ref{line:online_ar_start}--\ref{line:online_ar_end}
and \ref{line:online_diff_start}--\ref{line:online_diff_end}
describe profile generation using ProfileAR
and ProfileDiff, respectively.
Lines~\ref{line:online_coverage_start}--\ref{line:online_coverage_end}
perform coverage-based probing-set selection,
followed by probing and serving in
lines~\ref{line:online_probe_start}--\ref{line:online_probe_end}.

\begin{figure}[t]
\centering
\begin{tikzpicture}[
  font=\scriptsize,
  >={Latex[length=1.6mm,width=1.1mm]},
  link/.style={->,draw=black!75,line width=0.55pt,shorten >=0.2pt,shorten <=0.2pt},
  flowbox/.style={
    draw=black!72,fill=black!2,
    rounded corners=2pt,
    text width=2.17cm,align=center,
    inner xsep=2.5pt,inner ysep=3.2pt,
    minimum height=1.04cm
  }
]

\node[flowbox] (input) at (-2.75, 0)
  {Available information\\[-1pt]$\mathcal I$};
\node[flowbox] (encoder) at (0, 0)
  {History encoder\\[-1pt]$\mathbf M=f_\phi(\mathcal I)$};
\node[flowbox] (generator) at (2.75, 0)
  {Trained profile generator};

\node[flowbox] (coverage) at (-2.75,-1.35)
  {Greedy coverage\\[-1pt]$\mathcal P$, $|\mathcal P|=P$};
\node[flowbox] (sets) at (0,-1.35)
  {Acceptable sets\\[-1pt]$\{\widehat{\mathcal A}_r\}_{r=1}^{S}$};
\node[flowbox] (profiles) at (2.75,-1.35)
  {Sampled profiles\\[-1pt]$\{\mathbf x_0^{(r)}\}_{r=1}^{S}$};

\node[flowbox] (probe) at (-2.75,-2.70)
  {Probe selected\\[-1pt]beams $j\in\mathcal P$};
\node[flowbox] (feedback) at (0,-2.70)
  {Noisy, quantized feedback\\[-1pt]$\{\widetilde y_j\}_{j\in\mathcal P}$};
\node[flowbox] (serve) at (2.75,-2.70)
  {Serving beam\\[-1pt]$\widehat j$ from feedback};

\draw[link] (input.east) -- (encoder.west);
\draw[link] (encoder.east) -- (generator.west);
\draw[link] (generator.south) -- (profiles.north);
\draw[link] (profiles.west) -- (sets.east);
\draw[link] (sets.west) -- (coverage.east);
\draw[link] (coverage.south) -- (probe.north);
\draw[link] (probe.east) -- (feedback.west);
\draw[link] (feedback.east) -- (serve.west);
\end{tikzpicture}
\caption{Recovery workflow during deployment.}
\label{fig:online_workflow}
\end{figure}










\begin{algorithm}[t]
\caption{Post-interruption beam recovery procedure}
\label{alg:online}
\begin{algorithmic}[1]
\Require Available information $\mathcal{I}$;
trained generator
$\mathrm{G}\in\{\mathrm{ProfileAR},\mathrm{ProfileDiff}\}$;
$P$, $S$, $T$, $T_s$
\Ensure Probing set $\mathcal{P}$ and
serving beam $\widehat j$
\State Compute history memory
$\mathbf{M}=f_\phi(\mathcal{I})$
\If{$\mathrm{G}=\mathrm{ProfileAR}$}
    \State Initialize $S$ sequences
    with start tokens
    \label{line:online_ar_start}
    \For{$j=0,\ldots,K-1$}
        \State Sample the level of beam $j$
        in each sequence from its
        $T_s$-scaled conditional given
        the prefix and $\mathbf{M}$
    \EndFor
    \label{line:online_ar_end}
\Else
    \State Initialize $S$ fully masked profiles
    \label{line:online_diff_start}
    \For{$i=1,\ldots,T$}
        \State Predict $T_s$-scaled
        clean-level distributions
        for masked coordinates
        \State Apply \eqref{eq:reverse_transition}
        to reveal or retain masked
        coordinates, preserving previously
        revealed levels
    \EndFor
    \label{line:online_diff_end}
\EndIf
\State Construct
$\{\widehat{\mathcal{A}}_r\}_{r=1}^{S}$
using \eqref{eq:generated_sets}
\label{line:online_coverage_start}
\State Initialize $\mathcal{P}_0=\varnothing$
and mark all samples uncovered
\For{$i=1,\ldots,P$}
    \State Compute $d_i(j)$ for each
    unselected beam using
    \eqref{eq:greedy_gain}
    \State Select $j_i$ using
    \eqref{eq:greedy_selection},
    breaking ties by sample frequency
    and then beam-index order
    \State Update $\mathcal{P}_i$ and
    mark samples containing $j_i$ covered
\EndFor
\label{line:online_coverage_end}
\State Set $\mathcal{P}=\mathcal{P}_P$
and probe the selected beams
\label{line:online_probe_start}
\State Collect noisy, quantized feedback
and select $\widehat j$
using \eqref{eq:serve_rule}
\label{line:online_probe_end}
\end{algorithmic}
\end{algorithm}

\vspace{-4mm}
\subsection{Inference complexity}
\label{subsec:lowcomp}

ProfileDiff generates $S$ profiles through
$T$ denoising steps, processing
all $K$ beam coordinates in parallel each
step. ProfileAR instead generates $K$ coordinates
sequentially per profile, using key--value
caching. Thus, $T<K$ reduces the sequential
depth of diffusion generation, but does not
necessarily imply less computations.
For $N_{\mathrm{dec}}$ Transformer layers and
embedding dimension $d$, the approximate complexity of one
full-profile decoder evaluation is given by
\begin{equation}
\mathcal{O}\left(
N_{\mathrm{dec}}\left[
K^2d+KLd+(K+L)d^2
\right]+KdQ
\right),
\end{equation}
accounting for self-attention, cross-attention,
projections, and classification.
ProfileDiff performs $ST$ such evaluations,
whereas a cached ProfileAR rollout has
comparable asymptotic cost to one full-profile
evaluation per sample. Reusable history-memory
projections can be cached in both methods.
Greedy coverage selection requires
$\mathcal{O}(PSK)$ operations.


\vspace{-2mm}
\section{Experimental Setup and Dataset Construction}
\label{sec:sim_setup}
This section describes the simulated deployments, recovery-event generation, model configurations, and evaluation protocol used to assess the proposed approaches.

\vspace{-4mm}
\subsection{Deployments and recovery events}

We use DeepMIMO ray-traced channels at 28~GHz for Dallas, San Diego, and Boston5G~\cite{alkhateeb2019deepmimogenericdeeplearning}. {Dallas and San Diego each provide three independently evaluated single-cell deployments, while Boston contributes one deployment. Unless otherwise stated, the results are averaged over events pooled from these seven deployments. Trajectories follow covered streets within 100~m of the corresponding BS. The default array has $N=32$ and $K=64$ beams.} The trajectories evolve on a coverage-derived street graph with junctions, speeds of 4--14~m/s, random stops, and a sampling interval of $0.1$~s. To emphasize route uncertainty, the reference evaluation is constructed such that 80\% of interruption events encounter at least one junction, where different route choices can lead to distinct recovery conditions. Table~\ref{tab:parameters} summarizes the parameters.

\begin{figure}[t]
\centering
\includegraphics[width=\columnwidth]{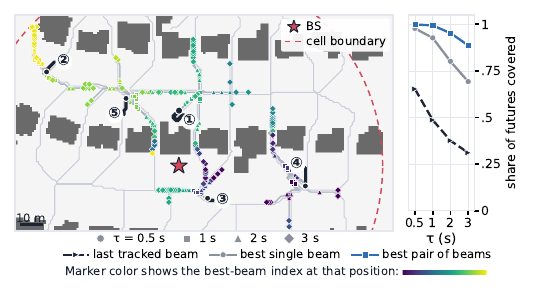}
\caption{Five recovery events in Dallas BS1 scenario. The left panel shows the map with buildings (dark gray). For each event, the black line is the tracked history ($L=10$ slots, 1~s), ending at the
dot where the link is lost, and the gray lines are 32 sampled futures. Markers show the recovery positions after $\tau=0.5$, 1, 2, and 3~s, and their color identifies the best beam at that position. The right panel shows the share of the futures covered, i.e., containing a beam within 3~dB of their best beam, versus $\tau$, averaged over 3000 simulated events reaching a junction, for the last tracked beam, the best single beam, and the best pair of beams of each event.}
\label{fig:scenarios}
\end{figure}

For each event, a history of served-beam indices and noisy, quantized SNR reports precedes an interruption of $\tau\in\{5,10,20,30\}$ slots. The reference histories use oracle beam tracking, which selects the strongest beam at each historical slot. Each event has eight independently evolved training futures or 32 test futures. Each deployment contains 30000 training, 2000 validation, and 4000 test events. {Events represent independent single-UE trajectories rather than simultaneous users, so no UE density is implied.} The default history length is $L=10$ and five training seeds are used. The acceptability margin is $\Delta_{\mathrm{acc}}=3$~dB. The proposed five-level quality profiles use thresholds of 3, 6, 10, and 15~dB relative to the strongest beam. Full-profile targets are generated for offline training and are not available at inference. {Fig.~\ref{fig:scenarios} illustrates how the same pre-interruption history can lead to distinct future positions and acceptable-beam sets near a junction.}

\begin{table}[t]
\centering\footnotesize
\caption{Default simulation and training configuration.}
\label{tab:parameters}
\setlength{\tabcolsep}{4pt}
\begin{tabular}{@{}p{0.38\columnwidth}p{0.55\columnwidth}@{}}
\toprule
\textbf{Parameter} & \textbf{Setting} \\
\midrule
Carrier; bandwidth & 28~GHz; 100~MHz \\
Transmit power; noise figure & $P_{\mathrm{tx}}=40$~dBm; $7$~dB \\
Array and codebook & {$N=32$; $K=64$} \\
Slot duration; cruise speed & 0.1~s; 4--14~m/s \\
Speed noise; stop rate & $1.5$~m/s/$\sqrt{\mathrm{s}}$; $0.03$~s$^{-1}$ \\
History; interruption & $L=10$; $\tau\in\{5,10,20,30\}$ slots \\
Feedback noise; quantization & $\sigma_v=1$~dB; $q=1$~dB \\
Train/validation/test events & 30000/2000/4000 per deployment \\
Futures per event & 8 training; 32 testing \\
Generated profiles & $S=64$  \\
Diffusion reverse steps & $T=64$\\
Transformer & $d=128$; 3 layers; 4 heads \\
Optimization & AdamW; $5\times10^{-4}$ learning rate; batch 512 \\
\bottomrule
\end{tabular}
\end{table}

\vspace{-4mm}
\subsection{Compared methods and evaluation protocol}\label{subsec:sim_baselines}
The generators use the same history interface, five-level target, encoder configuration, decoder width and depth, sample budget, and greedy coverage selector, and their parameters are trained separately. The primary baselines used throughout the numerical evaluation are described below. Additional baselines are introduced where relevant in the results sections.

\subsubsection{Tracking}
    Prioritizes beams near the last serving
    direction without predicting future quality.

\subsubsection{Exponential moving average (EMA) and upper confidence bound (UCB)}
    EMA ranks beams using exponentially weighted
    feedback, while UCB combines estimated quality
    with an observation-count-based exploration
    bonus~\cite{reinforceintro_sutton,UCB_ref}.

\subsubsection{Categorical predictors (TRM and ODE-LSTM)}
TRM predicts beam probabilities from historical
feedback, while ODE-LSTM incorporates
continuous-time recurrent dynamics
~\cite{ode_lstm_continuous_time_beam_pred}.
Both use soft targets obtained by applying
a softmax ($\tau_{\mathrm{lbl}}=2$~dB) to the
four strongest beams' SNRs, and zero
probability elsewhere. Unlike joint
generators, they model individual beam
preferences without joint
acceptability.

\subsubsection{Conditional variational autoencoder (CVAE)-profile}
    Generates joint discretized
    profiles conditioned on history, using a
    shared latent variable to model inter-beam
    dependencies and the same prediction target
    as the proposed methods.

The primary performance metric is the probe-failure probability,
estimated by averaging $F_{\mathrm p}$ over recovery events.
A probe failure occurs when the true SNR of every beam in
$\mathcal P$ falls more than $\Delta_{\mathrm{acc}}$ below the
optimum. For serving performance, we define
$D=s_{t_{\mathrm r}}^\star-s_{t_{\mathrm r},\widehat j}$ as the
SNR loss in dB of the served beam, where $\widehat j$ is selected
using the noisy, quantized probe reports. Thus, $D$ is computed
from the true SNRs, and it can exceed $\Delta_{\mathrm{acc}}$
even when the probing set contains an acceptable beam. We report
$\Pr(D>10~\mathrm{dB}\mid D>\Delta_{\mathrm{acc}})$, which we
refer to as the severe-loss probability in the following, and the
loss complementary cumulative distribution function (CCDF), $\Pr(D>x)$. For a metric with baseline value $M$,
we report the relative difference as
$100(M_{\mathrm{ProfileDiff}}/M-1)\%$, with negative values
favoring ProfileDiff.

\vspace{-2mm}
\section{Numerical Results}\label{sec:results}

Here, we evaluate the performance of the proposed methods.
All experiments were conducted using NVIDIA GH200 GPU.\footnote{Resources
were provided by CSC--IT, Finland,
through Roihu supercomputing infrastructure.
See \url{https://docs.csc.fi/computing/systems-roihu}.}

\vspace{-6mm}
\subsection{Beam-recovery performance}


\begin{figure}[t]
\centering
\includegraphics{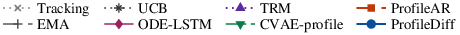}\\[2pt]
\includegraphics[width=0.95\columnwidth]{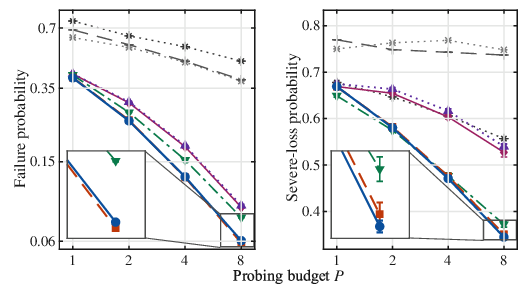}
\caption{{Failure probability (left) and severe-loss probability (right) versus $P$.}}
\label{fig:P}
\end{figure}

Fig.~\ref{fig:P} evaluates recovery performance
versus $P$. Increasing $P$ reduces
probe failure, with substantially
larger gains for ProfileAR and ProfileDiff as
more probes become available. This highlights
the benefit of modeling joint beam quality and selecting complementary probes.
{Both proposed methods perform comparably across the evaluated probing budgets.} Among events with serving loss above $\Delta_{\mathrm{acc}}$, the fraction exceeding 10~dB also decreases more for the proposed methods as $P$ increases. These results demonstrate the
benefits of joint-profile generation and
coverage-based selection in improving both
recovery reliability and the severity of
residual losses.

\begin{figure}[t]
\centering
\includegraphics{figs/legend_methods_col}\\[2pt]
\includegraphics[width=0.95\columnwidth]{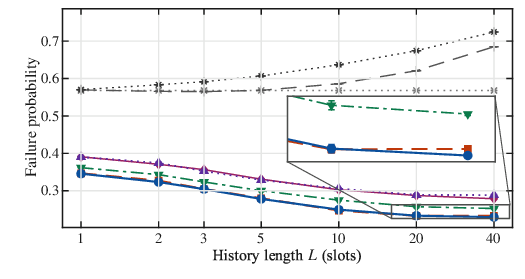}
\caption{{Failure probability versus history length.}}
\label{fig:L}
\end{figure}

Fig.~\ref{fig:L} examines the impact of
$L$ on recovery performance.
Learning-based methods generally benefit
from longer histories, with diminishing
returns, whereas Tracking remains unaffected
and EMA and UCB deteriorate due to outdated
observations. ProfileAR and ProfileDiff consistently achieve
the lowest failure, with their advantage
over categorical predictors increasing with
$L$. This demonstrates their ability
to exploit temporal information while capturing
uncertainty over recovery outcomes and CVAE-profile
remains less effective.

\begin{figure}[t]
\centering
\includegraphics{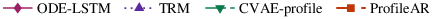}\\[2pt]
\includegraphics[width=0.95\columnwidth]{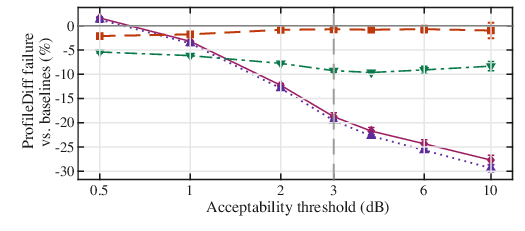}
\caption{{Relative failure of ProfileDiff against learning-based baselines versus the acceptability threshold. Negative values favor ProfileDiff.}}
\label{fig:delta}
\end{figure}

Fig.~\ref{fig:delta} examines sensitivity to
the acceptability threshold, varied only during
evaluation. Categorical predictors remain
competitive under strict thresholds, where
recovery primarily requires identifying the
strongest beam. As the acceptable quality margin
widens, ProfileDiff increasingly outperforms
these predictors, demonstrating the benefit
of modeling joint beam-quality profiles. ProfileDiff also consistently outperforms
CVAE-profile despite their common prediction
target, highlighting the importance of the
generative architecture. ProfileAR and
ProfileDiff perform comparably across most
thresholds. Overall, the results demonstrate
that joint-profile generation remains effective
beyond the training threshold, particularly
when multiple near-optimal beams are acceptable.

\vspace{-4mm}
\subsection{Robustness to observation and supervision quality}
\label{subsec:history_results}

\begin{figure}[t]
\centering
\includegraphics{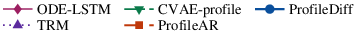}\\[2pt]
\includegraphics[width=0.46\columnwidth]{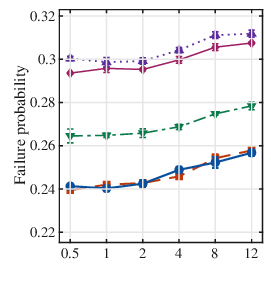}\hfill
\includegraphics[width=0.46\columnwidth]{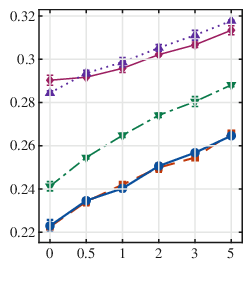}\\\vspace{-2mm}
\includegraphics[width=0.46\columnwidth]{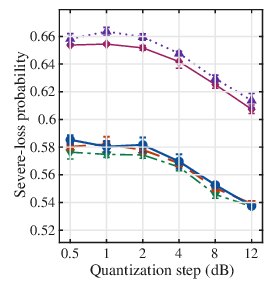}\hfill
\includegraphics[width=0.46\columnwidth]{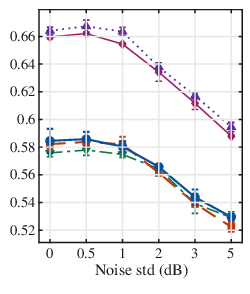}
\caption{Recovery performance versus feedback quality. Columns correspond to feedback quantization (left) and measurement noise (right), while rows show probe-failure probability (top) and severe-loss probability (bottom). Models are trained and evaluated at each feedback-quality setting.}
\label{fig:feedback}
\end{figure}

Fig.~\ref{fig:feedback} evaluates the impact of
historical feedback quality on recovery performance.
Probe failure increases gradually with coarser
quantization and stronger noise, with the latter
causing greater degradation over the evaluated
ranges. ProfileAR and ProfileDiff consistently
outperform the baselines, although their advantage
narrows slightly under stronger noise. CVAE-profile
also remains more effective than categorical
predictors. Although probe failure increases as feedback quality deteriorates, among events with serving loss above $\Delta_{\mathrm{acc}}$, the fraction exceeding 10~dB decreases slightly.
The profile-based generators consistently exhibit
less severe residual losses than categorical
predictors, further supporting their benefits.

\begin{figure}[t]
\centering
\includegraphics{figs/legend_learned_col}\\[2pt]
\includegraphics[width=\columnwidth]{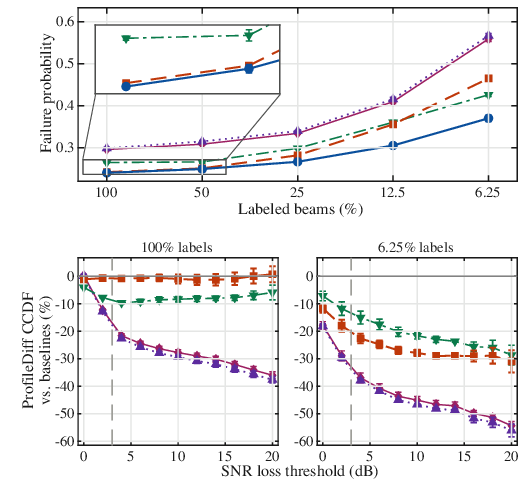}
\caption{{Recovery under incomplete training supervision. The top panel shows probe failure versus training-label availability. The lower panels show the relative difference in $\Pr(D>x)$ for ProfileDiff versus baselines as the SNR loss threshold $x$ varies, under complete (left) and 6.25\% (right) supervision. Dashed line marks the 3-dB acceptability threshold.}}
\label{fig:partial}
\end{figure}

Fig.~\ref{fig:partial} examines recovery under incomplete
training supervision. We randomly retain a subset of beam-quality
measurements for each training future and use the same subsets
across methods and training seeds. At evaluation, recovery is
scored relative to the strongest beam in the full codebook. Although all methods degrade as
labels become sparse, ProfileDiff exhibits
greater robustness than ProfileAR, with their
comparable performance under complete supervision
giving way to an increasing diffusion advantage.
CVAE-profile also degrades more gradually than
ProfileAR, surpassing it at the lowest label
fraction but remaining less effective than
ProfileDiff. These differences are consistent
with ProfileDiff's masked reconstruction of
observed levels, whereas ProfileAR must handle
incomplete autoregressive prefixes. Categorical
predictors additionally suffer from uncertain
best-beam labels.

{As seen in the lower panels of Fig.~\ref{fig:partial},} under complete supervision, ProfileAR
and ProfileDiff exhibit nearly identical loss
distributions, with their advantage over
categorical predictors increasing toward severe
losses. Under sparse supervision, ProfileDiff
outperforms all baselines across the evaluated
thresholds. {Its advantage grows toward the severe-loss tail in single-cell deployments.} Overall, diffusion's
robustness to incomplete supervision extends
beyond recovery success to reducing substantial
quality losses.

\vspace{-4mm}
\subsection{{Sampling budget}}
\label{subsec:sampling_budget}

\begin{figure}[t]
\centering
\includegraphics{figs/legend_learned_col}\\[2pt]
\includegraphics[width=0.9\columnwidth]{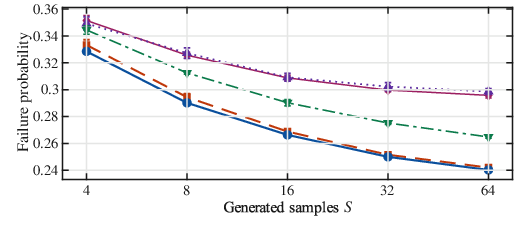}
\caption{{Failure probability versus the number of generated samples.}}
\label{fig:samples}
\end{figure}

{Fig.~\ref{fig:samples} examines the impact of
$S$ on recovery performance.
Increasing $S$ reduces probe failure, with diminishing returns
as empirical coverage becomes more representative
of possible recovery outcomes. ProfileAR and
ProfileDiff benefit particularly from additional
samples, as their joint profiles enable the
selection of complementary beams, unlike
categorical predictors that generate individual
beam indices. Notably, ProfileDiff exhibits
a slight advantage over ProfileAR at small
$S$, suggesting more effective
coverage with limited
sampling.}

\vspace{-4mm}
\subsection{{Generalization across deployments and conditions}}
\label{subsec:generalization_results}

{Here, we extend the evaluation to two
coordinated three-cell deployments, one each
in Dallas and San Diego, where the three BSs
previously evaluated independently are now
considered jointly. Trajectories follow streets
covered by at least one BS. With $C=3$ BSs and
$K=64$ beams per BS, each candidate is indexed
by $j=cK+k$, where $c\in\{0,\ldots,C-1\}$
and $k\in\mathcal K$ identify the BS and its
beam, respectively. The generators are trained
on joint profiles of $CK=192$ beam qualities,
with a probing budget $P$ shared across BSs.
The oracle and acceptable set are defined
over these pairs, while incomplete training
profiles use the strongest observed pair as
reference. This simplified coordinated setting
assumes equal probing costs and excludes
inter-cell interference and handover signaling
and delays. It aims to provide generalization
insights for a larger joint candidate space,
rather than capture the full complexity of
practical multi-cell operation.}

{Each historical observation identifies
the serving BS and beam, allowing handovers
during tracking and serving-cell changes at
recovery. We assume coordinated measurements
with equal probing costs, without modeling
inter-cell interference or handover signaling
and delays. The shift and ablation studies
below report pooled results, while
Section~\ref{subsec:multicell_results}
specifically examines coordinated multi-cell
recovery.}

\subsubsection{{Distribution shifts and adaptation}}

\begin{figure}[t]
\centering
\includegraphics[width=0.9\columnwidth]{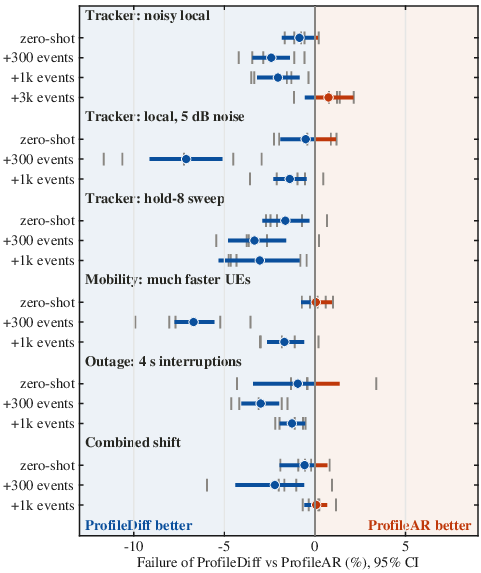}
\caption{Relative probe failure of ProfileDiff
versus ProfileAR under distribution shifts,
without adaptation and after fine-tuning on
300, 1000, or 3000 target events ($L=10$, $P=2$).
Shifts include noisy local tracking (1 or 5~dB
feedback noise), exhaustive sweeping every eighth
slot with beam holding in between, faster UEs
(14--24~m/s), longer interruptions (4~s),
and their combination. Bars denote 95\%
confidence intervals, dots pooled means,
and gray ticks deployment-specific means.}
\label{fig:regimes}
\end{figure}

Fig.~\ref{fig:regimes} evaluates the robustness
of generators to shifts in tracking
behavior, mobility, and interruption.
ProfileDiff generally achieves lower 
probe failure following limited target-domain
fine-tuning, with larger gains
under noisy local tracking and faster mobility.
This highlights the advantage of diffusion
in adapting to altered observation histories
and recovery conditions with limited data.
Although the performance gap narrows as more
adaptation data become available, the
results establish a clear benefit of ProfileDiff
in the data-limited adaptation regimes.

Table~\ref{tab:general} examines sensitivity to training-set size, transfer to unseen cities, joint training across deployments, and antenna-array and codebook configurations. All learned methods benefit from additional source training data, with ProfileAR and ProfileDiff consistently achieving the lowest failure rates. Their performance remains comparable across the evaluated source-data sizes, indicating that diffusion does not offer a systematic advantage simply because fewer source training events are available. Unseen-city zero-shot transfer is substantially weaker than in-domain evaluation for both models, and fine-tuning on target-city events markedly reduces failure. ProfileDiff has its clearest advantage with the smallest adaptation set, while the two methods converge as target data increase. 
In contrast, training a single model jointly
across deployments favors ProfileAR, potentially
reflecting the greater optimization demands of
diffusion when learning from heterogeneous
channel distributions. 
Both proposed methods maintain competitive
recovery performance across different array
sizes and codebook resolutions, demonstrating
the applicability of joint-profile recovery
across diverse configurations.

\begin{table}[t]
\centering\footnotesize
\caption{Probe failure across training-set sizes, sites and array/codebook configurations ($P=2$).}
\label{tab:general}
\setlength{\tabcolsep}{2.5pt}
\begin{tabular}{@{}lcccc@{}}
\toprule
\textbf{Condition (deployments)} & \textbf{TRM} &\textbf{ CVAE-profile} & \textbf{ProfileAR} & \textbf{ProfileDiff} \\
\midrule
In-domain (9) & 0.314 & 0.269 & \textbf{0.250} & 0.251 \\
\hline
\multicolumn{5}{@{}l}{\emph{{Source training events (5)}}} \\
\quad 1000 & 0.498 & 0.494 & \textbf{0.469} & 0.474 \\
\quad 3000 & 0.443 & 0.413 & \textbf{0.399} & 0.403 \\
\quad 10000 & 0.372 & 0.341 & 0.320 & \textbf{0.319} \\
\quad 30000 (default) & 0.326 & 0.286 & 0.267 & \textbf{0.266} \\
\hline
\multicolumn{5}{@{}l}{\emph{Unseen city (9)}} \\
\quad Zero-shot & 0.762 & 0.755 & 0.752 & \textbf{0.745} \\
\quad +300 target events & 0.519 & 0.507 & 0.509 & \textbf{0.489} \\
\quad +1000 target events & 0.452 & 0.449 & 0.406 & \textbf{0.403} \\
\quad +3000 target events & 0.400 & 0.371 & 0.344 & \textbf{0.341} \\
\hline
\multicolumn{5}{@{}l}{\emph{One shared model}} \\
\quad Single-cell (7) & 0.312 & 0.269 & \textbf{0.252} & 0.267 \\
\quad Three-cell (2) & 0.370 & 0.326 & \textbf{0.304} & 0.314 \\
\hline
\multicolumn{5}{@{}l}{\emph{Array size $N$/beams per BS $K$ (5)}} \\
\quad $8/16$ & 0.188 & 0.140 & \textbf{0.129} & 0.131 \\
\quad $32/32$ & 0.279 & 0.266 & 0.257 & \textbf{0.254} \\
\quad $32/128$ & 0.366 & 0.289 & \textbf{0.259} & \textbf{0.259} \\
\quad $64/128$ & 0.408 & 0.366 & 0.338 & \textbf{0.335} \\
\bottomrule
\end{tabular}
\vspace{2pt}
\parbox{\columnwidth}{\vspace{2pt}\scriptsize Numbers in parentheses denote deployments. Source models are trained from scratch on the stated number of training events. Unseen-city models are pretrained on other cities and fine-tuned on the stated target, while shared models pool the specified deployments. Each array/codebook configuration is trained and tested separately.}
\end{table}

\subsubsection{{Ablation studies}}
{The following ablations pool five deployments, one single-cell site in each city and two coordinated three-cell sites.}

\paragraph{{Generative representations and probing-set selection}}

Table~\ref{tab:selection} compares alternative
generative representations and probing-set
selection rules. ProfileAR and ProfileDiff
achieve the lowest failure probabilities,
outperforming both graded-profile and binary-set
generative baselines, and categorical
predictors. Notably, CVAE-profile is less
effective than CVAE-set, indicating that a
richer prediction target alone does not
guarantee better recovery and the generative
architecture also matters. Applying different selectors to the same samples isolates the
impact of coverage-based probing.
For models capturing joint beam acceptability,
greedy coverage consistently outperforms
frequency, marginal, and angular-diversity
selection, with larger gains at higher probing
budgets. By selecting beams that cover
previously uncovered channel outcomes, it
reduces redundant probing. In contrast,
independent-set and single-index predictors
benefit little from coverage because their
samples lack the joint structure.

\begin{table}[t]
\centering\footnotesize
\caption{Probe failure under alternative selection rules.}
\label{tab:selection}
\setlength{\tabcolsep}{2.2pt}
\begin{tabular}{@{}lcccccccc@{}}
\toprule
\textbf{Model} & \textbf{Cov$_2$} & \textbf{Freq$_2$} & \textbf{DPP$_2$} & \textbf{Marg$_2$}
      & \textbf{Cov$_4$} & \textbf{Freq$_4$} & \textbf{DPP$_4$} & \textbf{Marg$_4$} \\
\midrule
\multicolumn{9}{@{}l}{\emph{Dependent acceptable-beam sets}} \\
ProfileDiff  & \textbf{0.266} & 0.337 & 0.324 & 0.330 & \textbf{0.142} & 0.217 & 0.208 & 0.212 \\
ProfileAR    & \textbf{0.267} & 0.338 & 0.333 & 0.338 & \textbf{0.142} & 0.219 & 0.215 & 0.219 \\
CVAE-profile & \textbf{0.289} & 0.344 & 0.339 & 0.344 & \textbf{0.171} & 0.227 & 0.223 & 0.227 \\
AR-set       & \textbf{0.282} & 0.355 & 0.348 & 0.355 & \textbf{0.153} & 0.230 & 0.226 & 0.230 \\
CVAE-set     & \textbf{0.286} & 0.344 & 0.339 & 0.344 & \textbf{0.163} & 0.225 & 0.221 & 0.225 \\
\midrule
\multicolumn{9}{@{}l}{\emph{Independent per-beam sets}} \\
Direct-set & 0.337 & 0.336 & \textbf{0.331} & 0.336 & 0.216 & 0.215 & \textbf{0.212} & 0.216 \\
\midrule
\multicolumn{9}{@{}l}{\emph{Single-beam samples}} \\
IndexDiff   & 0.337 & 0.334 & \textbf{0.332} & 0.334 & 0.214 & 0.209 & \textbf{0.207} & 0.209 \\
TRM         & 0.326 & 0.323 & \textbf{0.316} & 0.321 & 0.201 & 0.198 & \textbf{0.194} & 0.198 \\
ODE-LSTM    & 0.327 & 0.324 & \textbf{0.318} & 0.323 & 0.203 & 0.199 & \textbf{0.194} & 0.198 \\
Direct-best & 0.317 & 0.315 & \textbf{0.308} & 0.310 & 0.186 & 0.184 & \textbf{0.176} & 0.179 \\
\bottomrule
\end{tabular}
\vspace{2pt}
\parbox{\columnwidth}{\vspace{2pt}\scriptsize
Subscripts indicate $P$. Cov.: greedy coverage;
Freq.: sample frequency; determinantal point process (DPP): quality-weighted
angular diversity; Marg.: marginal acceptance
probability ranking. AR-set generates up to eight
acceptable beams autoregressively; CVAE-set uses
binary targets instead of graded profiles.
Direct-set samples independent Bernoulli
acceptability variables; Direct-best samples
categorical best-beam predictions; IndexDiff
generates single beam indices through discrete
diffusion.}
\end{table}

\paragraph{{Profile representation and dependency modeling}}

Table~\ref{tab:abl_structure} examines the
representation, generation, and conditioning
components of ProfileDiff at two history lengths.
Binary and capped-set targets moderately increase
probe failure, whereas single-index generation
causes substantially larger degradation,
particularly at higher probing budgets.
Importantly, removing inter-beam attention
severely degrades recovery, highlighting the
role of the Transformer in capturing dependencies
across beam qualities. Replacing the Transformer
denoiser with a global MLP also increases
failure, although less substantially, indicating
that alternative architectures can capture
some of these dependencies. Similarly, removing
iterative reconstruction deteriorates
multi-beam recovery. These results
demonstrate the importance of joint-profile
representation, attention-based dependency, and iterative generation.

The conditioning ablations reveal that
interruption duration is particularly important,
while pooling historical observations has
relatively modest effects. Simpler beam-identity
representations, particularly fixed angular
and one-hot encodings, can match or outperform
the reference learned-plus-angular representation,
whereas a normalized scalar index generally
performs worse. The reference encoder was
retained as originally specified to avoid
retrospective selection based on the ablations.
Together with Table~\ref{tab:selection}, these
results identify joint outcome modeling and
coverage-based selection as the principal
sources of multi-probe recovery gains.

\begin{table}[t]
\centering\footnotesize
\caption{Structural ablations of ProfileDiff.}
\label{tab:abl_structure}
\setlength{\tabcolsep}{3pt}
\begin{tabular}{@{}lcccc@{}}
\toprule
\textbf{Configuration} & {$\textbf{({10,2})}$} & {$\textbf{(10,4)}$} &{ $\textbf{(40,2)}$} & {$\textbf{(40,4)}$} \\
\midrule
Full model failure probability & 0.266 & 0.142 & 0.254 & 0.136 \\
\midrule
\multicolumn{5}{@{}l}{\emph{Profile target}} \\
\quad Binary levels & $+5.6$ & $+6.7$ & $+6.6$ & $+10.1$ \\
\quad Set sequence (cap eight) & $+5.3$ & $+7.3$ & $+4.0$ & $+4.4$ \\
\quad Single best-beam index & $+26.8$ & $+50.3$ & $+27.7$ & $+49.2$ \\
\hline
\multicolumn{5}{@{}l}{\emph{Denoiser and generation}} \\
\quad Global MLP denoiser & $+4.8$ & $+7.9$ & $+3.9$ & $+6.2$ \\
\quad No inter-beam attention & $+26.0$ & $+51.9$ & $+27.0$ & $+52.4$ \\
\quad One reveal step ($T=1$) & $+24.0$ & $+48.0$ & $+26.2$ & $+50.5$ \\
\quad Separately trained one-step & $+26.0$ & $+51.6$ & $+27.1$ & $+52.7$ \\
\hline
\multicolumn{5}{@{}l}{\emph{Beam-identity tokens}} \\
\quad Learned embedding only & $+1.6$ & $+2.0$ & $+1.1$ & $+1.8$ \\
\quad Fixed angular code only & $-0.7^{\circ}$ & $-2.9^{\circ}$ & $-2.7$ & $-5.0$ \\
\quad One-hot beam index & $-1.7^{\circ}$ & $-4.0$ & $-3.3$ & $-5.4$ \\
\quad Normalized index $k/K$ & $+6.9$ & $+7.3$ & $+4.2^{\circ}$ & $+3.6^{\circ}$ \\
\hline
\multicolumn{5}{@{}l}{\emph{Conditioning and selection}} \\
\quad Interruption duration withheld & $+30.4$ & $+43.9$ & $+40.4$ & $+58.8$ \\
\quad Pooled history context & $-1.0^{\circ}$ & $-1.8^{\circ}$ & $-3.2$ & $-4.9$ \\
\quad Frequency instead of coverage & $+26.6$ & $+52.1$ & $+28.8$ & $+54.4$ \\
\bottomrule
\end{tabular}
\vspace{2pt}
\parbox{\columnwidth}{\vspace{2pt}\scriptsize
Columns indicate $(L,P)$. Ablations are reported
in relative changes (\%) averaged over five
deployments. $^{\circ}$ denotes inconsistent
signs across deployments.}
\end{table}

\paragraph{{Diffusion corruption process}}

Table~\ref{tab:abl_kernel} evaluates the
impact of alternative corruption kernels.
Independent absorbing masking achieves the
lowest probe failure and regret, indicating
that introducing spatially correlated masking
or ordinal quality-level transitions offers
no advantage in the evaluated settings.
Although uniform corruption slightly improves
exact-best-beam identification, it yields
higher probe failure, highlighting that
accurate prediction of the strongest beam
does not necessarily translate into better
multi-beam coverage.

\begin{table}[t]
\centering\footnotesize
\caption{ProfileDiff corruption-kernel ablations for $P=2$.}
\label{tab:abl_kernel}
\setlength{\tabcolsep}{3pt}
\begin{tabular}{@{}lccccc@{}}
\toprule
\textbf{Kernel} & \textbf{Failure} & \textbf{SNR} & \textbf{Regret} & \textbf{Miss} & \textbf{Cond.}\ \textbf{regret} \\
\midrule
\multicolumn{6}{@{}l}{\emph{Independent}} \\
\quad Absorbing & \textbf{0.266} & \textbf{28.48} & \textbf{4.18} & 0.508 & \textbf{8.15} \\
\midrule
\multicolumn{6}{@{}l}{\emph{Codebook correlated}} \\
\quad $\sigma=2$ & 0.271 & 28.43 & 4.23 & 0.511 & 8.20 \\
\quad $\sigma=4$ & 0.278 & 28.32 & 4.35 & 0.512 & 8.40 \\
\quad $\sigma=8$ & 0.284 & 28.25 & 4.42 & 0.514 & 8.50 \\
\midrule
\multicolumn{6}{@{}l}{\emph{Quality-level corruption}} \\
\quad Ordinal + mask & 0.274 & 28.37 & 4.29 & 0.508 & 8.36 \\
\quad Ordinal & 0.280 & 28.27 & 4.40 & 0.512 & 8.50 \\
\quad Uniform & 0.277 & 28.29 & 4.38 & \textbf{0.504} & 8.59 \\
\bottomrule
\end{tabular}
\vspace{2pt}
\parbox{\columnwidth}{\vspace{2pt}\scriptsize
SNR, regret, and conditional regret are in dB; Miss denotes the exact-best-beam miss probability. Regret is the oracle
SNR minus the served SNR; conditional regret is the oracle SNR minus the best probed beam SNR, averaged over the events in which the oracle beam is not probed. Independent absorbing masks
each beam independently. Correlated absorbing couples masking across neighboring beams, with $\sigma$ controlling the
correlation scale (in beams), while preserving marginal masking probabilities. Ordinal corruption perturbs quality
levels toward adjacent bins and ordinal + absorbing combines such perturbations with masking. Uniform perturbs without regard to ordinal proximity.}
\end{table}

\subsubsection{{Recovery in coordinated multi-cell deployments}}
\label{subsec:multicell_results}

\begin{figure}[t]
\centering
\includegraphics{figs/legend_methods_col}\\[2pt]
\includegraphics[width=0.95\columnwidth]{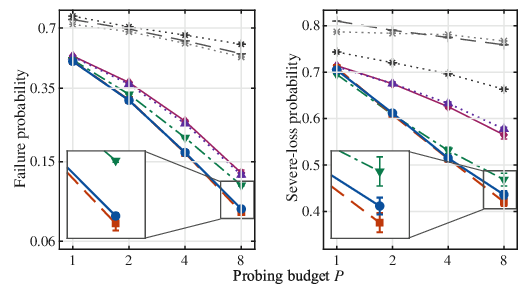}
\caption{{Three-cell recovery versus the shared probing budget $P$. The panels show probe failure (left) and severe-loss probability (right).}}
\label{fig:mc_recovery}
\end{figure}

{We now evaluate coordinated recovery across
the three BSs in Dallas and San Diego,
using a shared probing budget.
Unlike the preceding single-cell evaluations,
the candidate beams are jointly selected
across BSs.} {Fig.~\ref{fig:mc_recovery} isolates the two coordinated deployments. With complete supervision, ProfileAR and ProfileDiff perform comparably. As the shared probing budget increases, both reduce probe failure and the probability of severe serving loss more than the baselines.}

\begin{figure}[t]
\centering
\includegraphics{figs/legend_learned_col}\\[2pt]
\includegraphics[width=0.49\columnwidth]{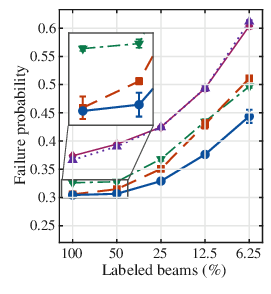}\hfill
\includegraphics[width=0.49\columnwidth]{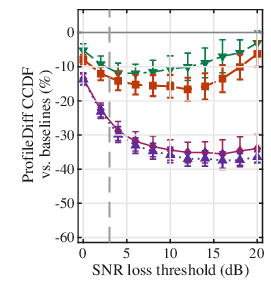}
\caption{{Three-cell recovery under incomplete training supervision. The left panel shows probe failure versus the fraction of labeled BS--beam pairs. The right panel shows the relative difference in $\Pr(D>x)$ for ProfileDiff versus each baseline as the loss threshold $x$ varies, with 6.25\% of labels.}}
\label{fig:mc_partial}
\end{figure}

{Fig.~\ref{fig:mc_partial} shows that the
advantage of ProfileDiff over ProfileAR
increases as fewer BS--beam pairs are labeled.
The greater sensitivity to incomplete supervision
here may stem from the larger
candidate space and additional uncertainty over
the serving BS, where sparse labels can miss
strong BS--beam alternatives. The loss-tail
curves also reveal a difference with single-cell settings.
With 6.25\% of labels, the reduction in tail
probability relative to ProfileAR and CVAE-profile
is largest at moderate losses and narrows at
the extreme tail. }

\vspace{-4mm}
\subsection{{Inference latency}}
\label{subsec:runtime_results}

\begin{figure}[t]
\centering
\includegraphics[width=0.85\columnwidth]{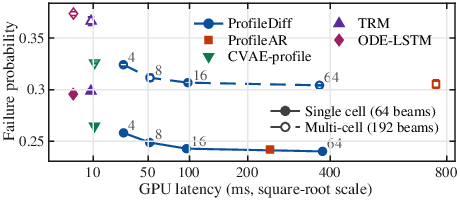}
\caption{{Probe failure probability versus GPU
inference latency in single-cell (filled markers)
and coordinated multi-cell (open markers)
deployments. ProfileDiff reverse steps are
indicated beside the markers.}}
\label{fig:latency}
\end{figure}

{Fig.~\ref{fig:latency} examines the tradeoff
between recovery performance and inference latency.
For ProfileDiff, increasing the number of reverse
steps improves recovery accuracy at the expense
of delay, with diminishing gains
beyond moderate step budgets. Reduced-step
diffusion approaches ProfileAR performance
with substantially lower latency,
particularly in multi-cell cases.
This advantage follows from parallel reconstruction
of beam-quality levels, whereas ProfileAR generates
them sequentially, increasing latency as the
candidate space grows. One-step predictors offer
the lowest latency but exhibit higher probe failure,
highlighting the flexibility of ProfileDiff in
balancing recovery accuracy and computational cost.}

{These measurements concern model inference
during recovery and exclude
reference-signal scheduling, probing, and feedback.
The simulation time step does not impose an NR
scheduling deadline and practical deployment requires
inference to finish before the probing opportunity
or its delay to be incorporated into the prediction
horizon. Accordingly, the reported latency
characterizes computational cost rather than
end-to-end recovery feasibility. Multi-UE scheduling
and alternative hardware platforms remain outside
the evaluation.}

\vspace{-2mm}
\section{Conclusions and Future Work}

This paper proposed two history-conditioned
generative approaches, ProfileAR and ProfileDiff,
for post-interruption beam recovery. By modeling
joint, graded beam-quality profiles and selecting
complementary probes through conditional coverage
maximization, both approaches improve recovery
over conventional predictors and generative
baselines. Extensive ablations establish the
importance of inter-beam dependencies, iterative
generation, and coverage-based selection. While both methods achieve comparable performance
under complete supervision, ProfileDiff exhibits
greater robustness to incomplete training labels
and outperforms ProfileAR under limited
adaptation to distribution shifts.

Future work may investigate training with
realistically collected sparse measurements,
accounting for measurement overhead, noise,
and nonrandom label availability. Further
directions include dynamic blockage, repeated
closed-loop recovery with online adaptation,
and end-to-end NR feasibility under practical
hardware and multi-UE scheduling constraints.

\vspace{-2mm}
\bibliographystyle{IEEEtran}
\bibliography{ref_abbv}

@article{Gen_wireless_1,
  author = {Khoramnejad, Fahime and others},
  title = {{Generative AI for the Optimization of Next-Generation Wireless Networks: Basics, State-of-the-Art, and Open Challenges}},
  journal = {IEEE Commun. Surveys Tuts.},
  year = {2025},
  pages = {3483--3525},
  doi = {10.1109/comst.2025.3535554},
  volume = {27},
  number = {6}
}

@article{Gen_wireless_2,
  author = {Van Huynh, Nguyen and others},
  title = {{Generative AI for Physical Layer Communications: A Survey}},
  journal = {IEEE Trans. Cogn. Commun. Netw.},
  year = {2024},
  volume = {10},
  number = {3},
  pages = {706--728},
  doi = {10.1109/tccn.2024.3384500}
}

@inproceedings{alkhateeb2019deepmimogenericdeeplearning,
  author = {Alkhateeb, Ahmed},
  title = {{DeepMIMO: A Generic Deep Learning Dataset for Millimeter Wave and Massive MIMO Applications}},
  year = {2019},
  url = {https://arxiv.org/abs/1902.06435},
  booktitle = {Proc. Inf. Theory Appl. Workshop (ITA)},
  pages = {1--8}
}

@book{reinforceintro_sutton,
  author = {Sutton, Richard S. and Barto, Andrew G.},
  title = {{Reinforcement Learning: An Introduction}},
  year = {1998},
  publisher = {MIT Press}
}

@article{giordaniBeamMgmt,
  author = {Giordani, Marco and others},
  title = {{A Tutorial on Beam Management for 3GPP NR at mmWave Frequencies}},
  journal = {IEEE Commun. Surveys Tuts.},
  year = {2019},
  volume = {21},
  number = {1},
  pages = {173--196},
  doi = {10.1109/comst.2018.2869411}
}

@techreport{ts38214,
  author = {{3GPP}},
  title = {{NR}; Physical Layer Procedures for Data},
  institution = {3GPP},
  type = {Tech. Spec.},
  number = {TS 38.214},
  note = {Version 19.5.0, Release 19},
  year = {2026},
  month = sep,
  url = {https://www.etsi.org/deliver/etsi_ts/138200_138299/138214/19.05.00_60/ts_138214v190500p.pdf}
}

@inproceedings{D3PM_main,
  author = {Austin, Jacob and others},
  title = {{Structured Denoising Diffusion Models in Discrete State-Spaces}},
  booktitle = {NeurIPS},
  volume = {34},
  pages = {17981--17993},
  publisher = {Curran Associates, Inc.},
  year = {2021}
}

@article{UCB_ref,
  author = {Auer, Peter and others},
  title = {{Finite-time Analysis of the Multiarmed Bandit Problem}},
  journal = {Mach. Learn.},
  year = {2002},
  volume = {47},
  number = {2--3},
  pages = {235--256},
  doi = {10.1023/a:1013689704352}
}

@article{mmwavesurvey2,
  author = {Wang, Xiong and others},
  title = {{Millimeter Wave Communication: A Comprehensive Survey}},
  journal = {IEEE Commun. Surveys Tuts.},
  year = {2018},
  volume = {20},
  number = {3},
  pages = {1616--1653},
  doi = {10.1109/comst.2018.2844322}
}

@article{rappapotmmwave,
  author = {Rappaport, Theodore S. and others},
  title = {{Millimeter Wave Mobile Communications for 5G Cellular: It Will Work!}},
  journal = {IEEE Access},
  year = {2013},
  volume = {1},
  pages = {335--349},
  doi = {10.1109/access.2013.2260813}
}

@article{heath_mmwavemimo,
  author = {Heath, Jr., Robert W. and others},
  title = {{An Overview of Signal Processing Techniques for Millimeter Wave MIMO Systems}},
  journal = {IEEE J. Sel. Topics Signal Process.},
  year = {2016},
  volume = {10},
  number = {3},
  pages = {436--453},
  doi = {10.1109/jstsp.2016.2523924}
}

@article{alrabeiah_sub6_to_mmwave_beam_blockage,
  author = {Alrabeiah, Muhammad and others},
  title = {{Deep Learning for mmWave Beam and Blockage Prediction Using Sub-6 GHz Channels}},
  journal = {IEEE Trans. Commun.},
  year = {2020},
  volume = {68},
  number = {9},
  pages = {5504--5518},
  doi = {10.1109/tcomm.2020.3003670}
}

@inproceedings{wang_situational_mmwave_beam_pred,
  author = {Wang, Yuyang and others},
  title = {{MmWave Beam Prediction with Situational Awareness: A Machine Learning Approach}},
  booktitle = {IEEE SPAWC},
  year = {2018},
  pages = {1--5},
  doi = {10.1109/spawc.2018.8445969}
}

@article{dualb_fusion_hetnet_beam_pred,
  author = {Ma, Ke and others},
  title = {{Deep Learning Assisted mmWave Beam Prediction for Heterogeneous Networks: A Dual-Band Fusion Approach}},
  journal = {IEEE Trans. Commun.},
  year = {2023},
  volume = {71},
  number = {1},
  pages = {115--130},
  doi = {10.1109/tcomm.2022.3222345}
}

@article{lowcomplex_ml_mmwave_beam_pred,
  author = {Khan, Muhammad Qurratulain and others},
  title = {{A Low-Complexity Machine Learning Design for mmWave Beam Prediction}},
  journal = {IEEE Wireless Commun. Lett.},
  year = {2024},
  volume = {13},
  number = {6},
  pages = {1551--1555},
  doi = {10.1109/lwc.2024.3381447}
}

@inproceedings{vision_position_multimodal_beam_pred,
  author = {Charan, Gouranga and others},
  title = {{Vision-Position Multi-Modal Beam Prediction Using Real Millimeter Wave Datasets}},
  booktitle = {IEEE WCNC},
  year = {2022},
  pages = {2727--2731},
  doi = {10.1109/wcnc51071.2022.9771835}
}

@inproceedings{lowfreq_prior_lstm_beam_pred,
  author = {Ma, Ke and others},
  title = {{Deep Learning Assisted mmWave Beam Prediction with Prior Low-frequency Information}},
  booktitle = {IEEE ICC},
  year = {2021},
  pages = {1--6},
  doi = {10.1109/icc42927.2021.9500788}
}

@article{lidar_future_beam_pred_v2i,
  author = {Jiang, Shuaifeng and others},
  title = {{LiDAR Aided Future Beam Prediction in Real-World Millimeter Wave V2I Communications}},
  journal = {IEEE Wireless Commun. Lett.},
  year = {2023},
  volume = {12},
  number = {2},
  pages = {212--216},
  doi = {10.1109/lwc.2022.3219409}
}

@article{multicell_multibeam_ae_lstm,
  author = {Shah, Syed Hashim Ali and others},
  title = {{Multi-Cell Multi-Beam Prediction Using Auto-Encoder LSTM for mmWave Systems}},
  journal = {IEEE Trans. Wireless Commun.},
  year = {2022},
  volume = {21},
  number = {12},
  pages = {10366--10380},
  doi = {10.1109/twc.2022.3183632}
}

@inproceedings{deepia_fast_initial_access,
  author = {Cousik, Tarun S. and others},
  title = {{Fast Initial Access with Deep Learning for Beam Prediction in 5G mmWave Networks}},
  booktitle = {MILCOM},
  year = {2021},
  pages = {664--669},
  doi = {10.1109/milcom52596.2021.9653011}
}

@article{ode_lstm_continuous_time_beam_pred,
  author = {Ma, Ke and others},
  title = {{Continuous-Time mmWave Beam Prediction With ODE-LSTM Learning Architecture}},
  journal = {IEEE Wireless Commun. Lett.},
  year = {2023},
  volume = {12},
  number = {1},
  pages = {187--191},
  doi = {10.1109/lwc.2022.3221159}
}

@article{multimodal_transformers_beam_pred,
  author = {Tian, Yu and others},
  title = {{Multimodal Transformers for Wireless Communications: A Case Study in Beam Prediction}},
  year = {2023},
  pages = {461--471},
  volume = {4},
  number = {3},
  doi = {10.52953/jwra8095},
  journal = {ITU J. Future Evol. Technol.}
}

@article{dl_beam_tracking_under_mobility,
  author = {Lim, Sun Hong and others},
  title = {{Deep Learning-Based Beam Tracking for Millimeter-Wave Communications Under Mobility}},
  journal = {IEEE Trans. Commun.},
  year = {2021},
  volume = {69},
  number = {11},
  pages = {7458--7469},
  doi = {10.1109/tcomm.2021.3107526}
}

@article{lstm_sub6_predictive_beam_tracking_v2i,
  author = {Zhao, Yao and others},
  title = {{LSTM-Based Predictive mmWave Beam Tracking via Sub-6 GHz Channels for V2I Communications}},
  journal = {IEEE Trans. Commun.},
  year = {2024},
  volume = {72},
  number = {10},
  pages = {6254--6270},
  doi = {10.1109/tcomm.2024.3395297}
}

@misc{beam_brainstorm_genssbf,
  author = {Zhou, Zihao and others},
  title = {{Beam-Brainstorm: A Generative Site-Specific Beamforming Approach}},
  year = {2026},
  eprint = {2601.02219},
  archiveprefix = {arXiv},
  primaryclass = {eess.SP},
  url = {https://arxiv.org/abs/2601.02219},
  note = {arXiv:2601.02219}
}

@article{beam_probe_pattern_deep,
  author = {Xue, Qiulin and others},
  title = {{Integrated Probing-Beam Pattern Learning and Beam Prediction for mmWave Massive MIMO}},
  journal = {IEEE Trans. Commun.},
  year = {2025},
  volume = {73},
  number = {8},
  pages = {6499--6513},
  doi = {10.1109/tcomm.2025.3538838}
}

@article{site_specific_probe,
  author = {Heng, Yuqiang and others},
  title = {{Learning Site-Specific Probing Beams for Fast mmWave Beam Alignment}},
  journal = {IEEE Trans. Wireless Commun.},
  year = {2022},
  volume = {21},
  number = {8},
  pages = {5785--5800},
  doi = {10.1109/twc.2022.3143121}
}

@article{Jsac_time_mmwave_beam,
  author = {Li, Qiaoyu and others},
  title = {{Machine Learning Based Time Domain Millimeter-Wave Beam Prediction for 5G-Advanced and Beyond: Design, Analysis, and Over-The-Air Experiments}},
  journal = {IEEE J. Sel. Areas Commun.},
  year = {2023},
  volume = {41},
  number = {6},
  pages = {1787--1809},
  doi = {10.1109/jsac.2023.3275613}
}

@article{beam_squint_ref,
  author = {Wang, Bolei and others},
  journal = {IEEE Trans. Signal Process.},
  title = {Beam Squint and Channel Estimation for Wideband {mmWave} Massive {MIMO-OFDM} Systems},
  year = {2019},
  volume = {67},
  number = {23},
  pages = {5893--5908},
  doi = {10.1109/tsp.2019.2949502}
}

@inproceedings{vaswani2017attention,
  author = {Vaswani, Ashish and others},
  title = {Attention Is All You Need},
  booktitle = {Advances in Neural Information Processing Systems},
  volume = {30},
  year = {2017},
  pages = {5998--6008}
}

@inproceedings{shi2024masked,
  author = {Shi, Jiaxin and others},
  title = {Simplified and Generalized Masked Diffusion for Discrete Data},
  booktitle = {Advances in Neural Information Processing Systems},
  volume = {37},
  year = {2024},
  pages = {103131--103167}
}

@article{nemhauser1978analysis,
  author = {Nemhauser, George L. and others},
  title = {An Analysis of Approximations for Maximizing Submodular Set Functions---{I}},
  journal = {Math. Program.},
  volume = {14},
  pages = {265--294},
  year = {1978},
  doi = {10.1007/bf01588971},
  number = {1}
}

@article{pei2025latent,
  author = {Pei, Jianhua and others},
  title = {{Latent Diffusion Model-Enabled Low-Latency Semantic Communication in the Presence of Semantic Ambiguities and Wireless Channel Noises}},
  journal = {IEEE Trans. Wireless Commun.},
  volume = {24},
  number = {5},
  pages = {4055--4072},
  year = {2025},
  doi = {10.1109/twc.2025.3535714}
}

@article{kim2026beyond,
  author = {Kim, Wonjung and others},
  title = {{Beyond Prompts: Preserving Semantics in Diffusion-Based Semantic Communication}},
  journal = {IEEE Wireless Commun. Lett.},
  volume = {15},
  pages = {2209--2213},
  year = {2026},
  doi = {10.1109/lwc.2026.3673359}
}

@misc{wang2026beamstep,
  author = {Wang, Ziyue and others},
  title = {{Beam-STeP: Generative Site-Specific Beam Prediction with Spatio-Temporal Priors}},
  year = {2026},
  eprint = {2610.20282},
  archiveprefix = {arXiv},
  primaryclass = {eess.SP},
  url = {https://arxiv.org/abs/2610.20282},
  note = {arXiv:2610.20282}
}

@misc{wang2026radiodiffv2,
  author = {Wang, Xiucheng and others},
  title = {{RadioDiff-v2: Generative Angular Radio Maps for Multi-Beam Selection and Localization}},
  year = {2026},
  eprint = {2607.08045},
  archiveprefix = {arXiv},
  primaryclass = {cs.IT},
  url = {https://arxiv.org/abs/2607.08045},
  note = {arXiv:2607.08045}
}

@inproceedings{othman2026diffusionpriors,
  author = {Othman, Esraa Fahmy and others},
  title = {{Diffusion-Based Generative Priors for Efficient Beam Alignment in Directional Networks}},
  booktitle = {Proc. IEEE Int. Conf. Commun. Workshops (ICC Workshops)},
  year = {2026},
  doi = {10.1109/iccworkshops63917.2026.11586354},
  pages = {1--6}
}

@article{zhang2023limitedsampling,
  author = {Zhang, Jianjun and others},
  title = {{Beam Training and Tracking With Limited Sampling Sets: Exploiting Environment Priors}},
  journal = {IEEE Trans. Commun.},
  volume = {71},
  number = {5},
  pages = {3008--3023},
  year = {2023},
  doi = {10.1109/tcomm.2023.3239927}
}

@article{bonfante2022predictiverecovery,
  author = {Bonfante, Andrea and others},
  title = {{Performance of Predictive Indoor mmWave Networks With Dynamic Blockers}},
  journal = {IEEE Trans. Cogn. Commun. Netw.},
  volume = {8},
  number = {2},
  pages = {812--827},
  year = {2022},
  doi = {10.1109/tccn.2021.3118459}
}

@misc{tang2026decoupledsensing,
  author = {Tang, Yao and others},
  title = {{Generative Site-Specific Beamforming for UPAs via Decoupled Channel Sensing}},
  year = {2026},
  eprint = {2606.22522},
  archiveprefix = {arXiv},
  primaryclass = {eess.SP},
  url = {https://arxiv.org/abs/2606.22522},
  note = {arXiv:2606.22522}
}

@inproceedings{parmar2018image,
  title = {Image transformer},
  author = {Parmar, Niki and others},
  booktitle = {Proc. ICML},
  pages = {4055--4064},
  year = {2018},
  volume = {80},
  publisher = {PMLR}
}

@techreport{alves20256gresiliencewhite,
  title = {{6G Resilience White Paper}},
  author = {Alves, Hirley and others},
  year = {2025},
  url = {https://oulurepo.oulu.fi/handle/10024/59226},
  institution = {Univ. Oulu},
  type = {6G Research Visions},
  number = {15}
}
\end{document}